\documentclass[aps,prd,psfig,amsmath,amsfonts,groupedaddress,eqsecnum]{revtex4}

\newcommand{\fourvec}[1]{\mbox{\boldmath $\mathsf{#1}$}}
\begin{document}

\input psfig.sty


\title{A Post-Newtonian diagnostic of  quasi-equilibrium binary
configurations of compact objects}


\author{Thierry Mora }
\email[]{tmora@clipper.ens.fr}
\affiliation{\'Ecole Normale Sup\'erieure, 
\\45 rue D'Ulm, 75230 Paris Cedex 05, France \footnote{Present address}}
\affiliation{McDonnell Center for the Space Sciences, Department of
Physics, \\Washington University, St. Louis, Missouri 63130}
\author{Clifford M. Will}
\email[]{cmw@wuphys.wustl.edu}
\homepage[]{wugrav.wustl.edu/people/CMW}
\affiliation{Groupe Gravitation Relativiste et Cosmologie (GR$\varepsilon$CO)\\
Institut d'Astrophysique, 98 bis Boulevard Arago, 75014 Paris, France}
\affiliation{McDonnell Center for the Space Sciences, Department of
Physics, \\Washington University, St. Louis, Missouri 63130
\footnote{Permanent address}}


\date{\today}

\begin{abstract}
Using equations of motion accurate to the third post-Newtonian (3PN)
order ($O(v/c)^6$ beyond Newtonian gravity), we derive
expressions for the total energy $E$ and angular momentum $J$ of the orbits of
compact binary systems (black holes or neutron stars)
for arbitrary orbital eccentricity.  
We also incorporate finite-size contributions
such as spin-orbit and spin-spin coupling, and rotational and tidal distortions,
calculated to the lowest order of approximation, but we exclude the
effects of gravitational radiation damping.  
We describe how these formulae may be used as an accurate diagnostic of the
physical content of 
quasi-equilibrium
configurations of compact binary systems of black
holes and neutron stars generated using numerical relativity.
As an example, we show that quasi-equilibrium configurations of corotating
neutron stars recently reported by Miller {\it et al.} can be fit by our
diagnostic to better
than one per cent with a circular orbit and with physically reasonable tidal
coefficients.
\end{abstract}


\maketitle

\section{Introduction and Summary }
\label{intro}
%
%
%
The late stage of inspiral of binary systems of neutron stars or black
holes is of great current interest, both as a challenge for numerical
relativity, and as a possible source of gravitational waves detectable
by laser interferometric antennas.  Because this stage, corresponding
to the final few orbits and ultimate merger of the two objects into
one, is highly dynamical and involves strong gravitational fields, it
must be handled by numerical relativity, which attempts to solve the
full Einstein equations on computers 
(see Refs. \cite{cook,seidel,baumshapiro} for
reviews).

The early stage of inspiral can be handled accurately using
post-Newtonian techniques, which involve an expansion of solutions of
Einstein's equations in powers of $\epsilon \sim (v/c)^2 \sim
Gm/rc^2$, where $v$, $m$, and $r$ are the typical velocity, mass and
separation in the system, respectively.  By expanding to very high
powers of $\epsilon$, one can derive
increasingly accurate formulae to
describe both the orbital motion and the gravitational waveform.
Currently, results for the orbital motion
accurate through 3.5PN order ($O(\epsilon^{7/2})$
beyond Newtonian gravity) are known 
\cite{jaraschafer98,jaraschafer99,djs00,djs01,bf00,bf01,abf01,
bdgef03,futamase01,itohfuta03,dire2}.

An important issue in understanding the full inspiral of compact
binaries is how to connect the PN regime to the numerical regime.
This is a non-trivial issue because the PN approximation gets worse
the smaller the separation between the bodies.  
On the other hand,
because of limited computational resources, numerical simulations
cannot always be started with separations sufficiently large to
overlap the PN regime where it is believed to be reliable.  
This has
given rise to the so-called Intermediate Binary Black Hole (IBBH)
problem \cite{ibbh}, for example, which seeks new 
techniques or insights to
attempt to bridge the gap between the end of confidence in PN methods
and the beginning of realistic numerical simulations.  On the other
hand, if it can be demonstrated that PN approximations converge
sufficiently rapidly, especially for comparable-mass binary systems,
then IBBH techniques may not be needed.  Blanchet \cite{luc02,luchopkins}
has recently argued that, for comparable-mass systems, the PN
approximation seems to be more accurate than might be expected based on
experience with the test-body limit.  For binary
neutron stars, this is less of an issue, because neutron stars are
much larger objects, so the numerical simulations necessarily commence at
larger separations, where PN methods are presumably more reliable.

Numerical simulations of compact binary inspiral start with a solution
of the initial value equations of Einstein's theory; these
provide the initial data for the evolution equations (some 
initial-data models
\cite{ggb02} solve in addition
one of the six dynamical field equations).  The initial
state is assumed to consist of two compact objects (neutron stars or
black holes) in an initially circular orbit.  
For stellar-mass systems
that have evolved in isolation for eons, gravitational radiation is
expected to leave the orbit in an accurately circular state, apart
from the adiabatic inspiral induced by the loss of orbital energy;
that inspiral is ignored in the 
initial-data models.
(Miller has analysed the consequences of this particular assumption
\cite{miller03}).

The circular-orbit condition is
imposed by demanding that $dr/dt = 0$ initially, where $r$ is a
measure of the orbital separation.  One way to achieve this 
is to require that the system 
have an initial ``helical Killing vector'' (HKV), which
corresponds to a kind of rigid rotation of the binary system.  
Some 
initial-data models assume that the objects
are co-rotating, a condition which is astrophysically unlikely,
albeit computationally advantageous, while others
assume that the bodies are irrotational, {\em i.e.}
non-rotating in an inertial frame.
To simplify the problem, an approximation for the spatial metric is
generally made; one is the assumption of 
conformal flatness, an approximation that is known to be invalid in full
general relativity.  This approximation is usually justified by the
neglect of radiation reaction in the initial state.  Other approximations,
derived from post-Newtonian theory, or from sums of Kerr geometries, have
also been used.
For black hole binaries, suitable horizon boundary conditions must be
imposed, while for neutron star binaries,  equations of
hydrostationary equilibrium and
an equation of state must be provided.  

One important product of these initial value solutions is a
relationship between the energy $E$ and angular momentum $J$
of the system as measured at infinity, and the orbital
frequency $\Omega$.  
The energy could be the total energy as measured
at infinity, consisting of the
masses of the two stars plus the orbital energy, or it could be the
total energy less the energy of the same two stars in isolation.  
The latter quantity would be a
measure of the orbital binding energy.   
As all  quantities 
are well-defined and gauge invariant, they are useful
variables for making comparisons with PN methods.  

We have developed  formulae for $E(\Omega)$ and $J(\Omega)$
using PN methods.  Our analytic formulae include point-mass terms
through 3PN order, but ignore radiation reaction.  
They also include rotational energy and
spin-orbit and spin-spin terms for the case in which the bodies are rotating.  
They  further include
a Newtonian calculation of 
the effects of tidal and rotational distortions, applicable to stars of
arbitrary density distribution, expressed in terms of so-called ``apsidal
constants'' ({\it i.e.} we do not restrict attention to homogeneous ellipsoids
\cite{wisemanlai}),
and including effects at quadrupole and octupole order.   
We verify that, for black holes, tidal
effects can be ignored, while for neutron-star binaries, they must be
included.  
In contrast to previous work \cite{luc02,cook94,pfeiffer,dgg02}, our
formulae apply to general eccentric orbits, not just to circular orbits.

In an earlier paper \cite{morawill}, we compared this formula with
HKV numerical solutions for corotating binary black holes obtained by
Grandcl\'ement et al. \cite{ggb02}, 
for the regime where the black holes are 
separated from the location of the innermost circular orbit by a
factor of around two, where PN
results might be expected to work well ($Gm/rc^2 \sim 0.1$). 
We found that
when we assumed circular PN orbits, our 3PN formulae for 
$E(\Omega)$ and $J(\Omega)$ agreed to within 0.5 \% 
with other PN methods, including our own formulae truncated at 2PN order, and
3PN formulae derived
using resummation or Pad\'e techniques.  However all PN methods 
consistently and systematically underestimated the binding energy and
overestimated the angular momentum, compared
to
the values derived from the numerical HKV 
initial-data models, by amounts that were up to 10 times larger
than the spread among the PN methods.
But when we relaxed the assumption of a circular orbit and demanded only
that $dr/dt =0$, our PN formula could be made to agree 
extremely well with the numerical 
data by assuming that the system being simulated is initially at the
apocenter of a slightly eccentric orbit.  For values of $Gm \Omega/c^3 $
ranging from 0.03 to 0.06, corresponding to orbital $v/c$ between 0.3 and 0.4, 
or orbital separation between 10 and 6 $Gm/c^2$, 
nearly perfect agreement with the binding energy and the angular
momentum could be obtained
with eccentricities that
range from 0.03 to 0.05.

The concordance within fractions of a percent
between the various 2PN, 3PN and resummation PN results
matches expectation, since
$(Gm/rc^2)^3
\sim 10^{-3}$.  Presuming that all relevant physical effects have been
included, we argued that the PN results in this range of
$Gm\Omega/c^3$ are  robust.
We suggested the  possibility that the approximations made in most
numerical
initial-data models could 
lead to an apparent eccentricity in what was expected to be a
quasi-circular orbit.     
At present, however, the discrepancy between the two approaches
can only be considered a hint of possible
eccentricity,
because the results of \cite{ggb02} did not include
quantitative error bars for the variables $E(\Omega)$ and $J(\Omega)$.

These results motivate us to propose a ``post-Newtonian
diagnostic'', a tool that can be used 
to extract physical information from numerical simulations,
and that may also be an aid
to guide some of the assumptions and approximations inherent
in
numerical initial data computations toward those that lead to the
desired
physical configuration, such as a true quasi-circular orbit.

In this paper we provide the physical assumptions, mathematical details, and
justifications for the approximations that underly this proposed diagnostic
tool.  We give the detailed foundations for the analysis carried
out in \cite{morawill} for black-hole binary systems, and also extend that
work to the case of neutron-star systems by including tidal effects.
As an application of our diagnostic to neutron-star systems, we analyse
recent numerical models of
quasi-equilibrium orbits of neutron stars by Miller {\it et al.}
\cite{mgs03}.  In contrast to the black-hole case, 
we find that the orbital energy in the neutron-star initial-data models of
\cite{mgs03} can be
fit 
to better than one percent, and importantly, within the error bars
provided in \cite{mgs03}, 
using circular orbits with physically
reasonable tidal parameters appropriate to the 
``$\Gamma=2$'' equation of state used in that
numerical work.  The results illustrate the robustness of the PN
approximation well into the strongly relativistic regime of compact binaries,
especially when augmented with physically movitated finite-size effects.
Application of this PN diagnostic to other numerical models will be
subject of future papers.

The remainder of this paper provides the details underlying these
conclusions.
In Sec. \ref{3pnsolution}, we solve the post-Newtonian  equations of motion
calculated to third post-Newtonian 
(3PN) order, for general
eccentric orbits.  Neglecting radiation reaction effects, we 
then express the total conserved orbital 
energy and angular momentum in terms of a
pair of ``covariant'' orbit elements $e$ (eccentricity) and $\zeta$ (related
to the semi-latus rectum).  In Sec. \ref{finitesize}, we calculate the effects
of finite size in binary systems with bodies whose spin axes are
perpendicular to the orbital plane.  These include tidal and rotational
distortions, spin-orbit terms and spin-spin terms.  In Sec.
\ref{pndiagnostic},
we analyse our diagnostic quantitatively, and apply it to co-rotating,
equal-mass binaries of black holes and of neutron stars.  
Two appendices provide the detailed
derivations of the expressions for the tidal and rotational distortion
included 
in our diagnostic: Appendix
\ref{appa} uses Newtonian gravity to solve the general problem of the
equilibrium configurations of gravitating bodies disturbed by an external
force, paralleling the treatment in the classic monographs of Kopal
\cite{kopal1,kopal2}, and Appendix
\ref{appb} specializes the results to linear perturbations caused by
rotational and tidal disturbances.  

\section{Energy and angular momentum for ``point'' masses to 3PN order}
\label{3pnsolution}

\subsection{Orbits at the turning point in post-Newtonian gravity}
\label{turningpoint}

Since our ultimate focus will be on orbits that are possibly eccentric, but
that momentarily have $\dot r = 0$, it will be useful to review the
characteristics of orbits at turning points in Newtonian theory.
In Newtonian gravity, the orbit of a pair of point masses may be
described by the set of equations
\begin{eqnarray}
p/r &=& 1+e \cos(\phi - \omega)  \,, 
\nonumber \\
r^2 \Omega &\equiv& r^2 d \phi /dt = (mp)^{1/2} \,,
\nonumber \\
E &=& \mu ( {\dot r}^2 + r^2 \Omega^2 )/2 - \mu m /r \,,
\nonumber \\
J &=& \mu |{\bf x} \times {\bf v}| \,,
\label{newton}
\end{eqnarray}
where $p = a(1-e^2)$ is the semi-latus rectum ($a$ is the semi-major
axis), $\omega$ is the angle of pericenter, $m=m_1 + m_2$ is the total
mass, $\mu = m_1m_2/m$ is the reduced mass, and $E$ and $J$ are the total
orbital energy and angular momentum, respectively 
(henceforth we use units in which $G=c=1$).  A circular orbit
corresponds to $e=0$, with $r=a=$ constant, $\Omega^2 = m/a^3$,
$E/\mu = a^2 \Omega^2/2 - m/a = -(m\Omega)^{2/3}/2$, and $J/\mu =
\sqrt{ma} = (m/\Omega)^{1/3}$.  However, if
we  demand only that the orbit be at apocenter, so that $\dot r = 0$
only, we have $\phi = \omega + \pi$, $r = p/(1-e)$, $\Omega^2 = (m/p^3)(1-e)^4$,
so that, in terms of  $\Omega_a$, the angular velocity at apocenter,
\begin{eqnarray}
E/\mu &=&  - \frac{1}{2} (1-e^2) \left [ \frac{m\Omega_a}{(1-e)^2} 
\right ] ^{2/3} \,, 
\nonumber \\ 
J/\mu m&=& \left [ \frac{m\Omega_a}{(1-e)^2} \right ] ^{-1/3}.
\label{EJapocenter}
\end{eqnarray}
To obtain 
expressions in terms of $\Omega_p$, the angular velocity at
pericenter, one makes the replacements
$\Omega_a \to \Omega_p$ and $e \to -e$ in Eqs. (\ref{EJapocenter}). 

However, at higher PN orders, 
neither the orbital eccentricity ${e}$ nor
the semi-latus rectum ${p}$ 
is uniquely or invariantly defined.  One definition of eccentricity
used by Lincoln and Will \cite{lincwill} in their analysis of orbits at
2.5PN order was that of a Newtonian orbit
momentarily tangent to the true orbit (the ``osculating''
eccentricity);  it 
had the unusual property that it did not tend to zero
for a circular PN orbit, but tended toward a constant value of order
$m/p$, while the rate of pericenter advance 
approached the same rate of rotation
as the orbit itself.  In this language, the true
orbit was a non-circular orbit at
perpetual periastron, thereby maintaining a constant separation $r$.
In an effort to avoid this anomaly,
other authors \cite{ds88} adopted a ``quasi-Keplerian'' parametrization,
which defined multiple
``eccentricities'' to encapsulate different aspects of non-circular
orbits at PN order.    

In an effort to find a parametrization of non-circular PN orbits that will
be useful in comparing with numerical models, we \cite{morawill}
proposed
an alternative
measure of eccentricity and semi-latus
rectum according to:
\begin{eqnarray}
e &\equiv& \frac{ \sqrt{\Omega_p} - \sqrt{\Omega_a} } 
	{ \sqrt{\Omega_p} + \sqrt{\Omega_a} } \,,
\nonumber \\
\zeta \equiv
\frac{m}{p} &\equiv& \left ( \frac{\sqrt{m\Omega_p} +
\sqrt{m\Omega_a}}{2} \right )^{4/3} \,,
	\label{ezeta}
\end{eqnarray}
where $\Omega_p$ is the value of $\Omega$ where it passes through a
local maximum (pericenter), and $\Omega_a$ is 
the value of $\Omega$ where it passes
through the {\it next}
local minimum (apocenter).  

These definitions have the following virtues: (1) they reduce precisely to
the normal eccentricity $e$ and semi-latus rectum $p$
in the Newtonian limit, as can be verified
from Eqs. (\ref{newton}); 
(2) they are constant in the absence of radiation reaction;
(3) they are somewhat more directly connected to measurable quantities, 
since $\Omega$ is the
angular velocity as seen from infinity (eg. as measured in the
gravitational-wave signal) and one calculates only maximum
and minimum values, without concern for the coordinate location in the
orbit; and (4) they are straightforward to calculate in a numerical
model of orbits without resorting to complicated definitions of
``distance'' between bodies.  

They have the defect that, when
radiation reaction is included, they are not local, continuously evolving
variables, but rather are some kind of orbit-averaged quantities (for
this reason, they may not be as ``covariant'' as they seem -- see Sec.
\ref{gauge} below).  Nevertheless,
when an eccentric orbit decays and circularizes 
under radiation reaction the definition of $e$
has the virtue that it tends naturally to
zero when the orbital frequency turns from ocillatory behavior to
monotonically increasing behavior (i.e. the maxima and minima merge).

By virtue of these definitions, $\zeta$ has the further property that
\begin{equation}
\zeta = \left ( \frac{m\Omega_p}{(1+e)^2} \right )^{2/3} 
	= \left ( \frac{m\Omega_a}{(1-e)^2} \right )^{2/3} \,.
	\label{zetaproperty}
\end{equation}

We will derive expressions for orbital energy and angular momentum in terms
of these parameters $e$ and $\zeta$; for comparision with numerical
models of quasi-equilibrium parametrized 
in terms of $\Omega$ at $\dot r =0$ ($\Omega_a$ or $\Omega_p$), one can
simply substitute for $\zeta$ from Eq. (\ref{zetaproperty}).  
In this section we will focus on 3PN expressions for point masses; in the
next section, we will incorporate effects due to rotation and finite size.

\subsection{3PN equations of motion}

We use the standard form of the equations of motion, written 
in a ``Newtonian-like'' manner.  The acceleration of body 1 is given
schematically by
\begin{eqnarray}
\mathbf{a}_{1}  =  \frac{d^{2} \mathbf{x}_{1}}{d t^{2}}
& =& \frac{m_2}{r^2}\left \{
\mathbf{n}[-1+(PN)+(2PN)+(2.5PN)
\right . 
\nonumber \\
&& \left .
\qquad +(3PN)+(3.5PN)
+\cdots] \right. 
\nonumber\\ 
&& \left. 
\qquad +\mathbf{v}[(PN)+(2PN)+(2.5PN)
\right .
\nonumber \\
&& 
\left . 
\qquad +(3PN)+(3.5PN)
+\cdots]\right \},
\end{eqnarray}
where $\mathbf{x}_{a}$ and $m_{a}$ denote the position and the 
mass of the body $a$, 
$r$ is the separation 
between the two bodies, $\mathbf{n} = (\mathbf{x}_1 - \mathbf{x}_2)/r$ 
is the unit vector from 2 to 1, and
$\mathbf{v} = \mathbf{v}_1 - \mathbf{v}_2$ the relative velocity.
The equation for body 2 is obtained by making the replacement $1
\rightleftharpoons 2$.
The notation $nPN$ represents the $n^{th}$ post-Newtonian 
correction to Newtonian gravity. These equations are valid 
only for point-like, non-spinning bodies.  

Post-Newtonian terms $nPN$ include even (integer) and 
odd (half-odd integer, such as 2.5PN, or 5/2 PN) 
orders. Even terms are conservative, 
in the sense that the equations of motion admit conserved quantities
such as energy and angular momentum. 
Odd terms correspond to gravitational radiation reaction, and 
therefore are not conservative. In particular, 
they will cause the orbit to shrink, and 
the eccentricity to decrease.

We convert the two-body problem to 
an effective one-body problem.  For this purpose we choose the 
origin to be at the center of mass of the system, which is defined 
by an integral of the motion (a conserved quantity to the 3PN order of
approximation to which we will be working). 
We then change all variables to 
the relative coordinates $\mathbf{x}=\mathbf{x}_{1}-\mathbf{x}_{2}$ 
using relations of the type
\begin{eqnarray}
\mathbf{x}_{1} & = & [m_{2}/m+ (\eta \delta m/2m)(v^2-m/r) 
 + (2PN)
+\cdots]\mathbf{x}\,,\nonumber\\
\mathbf{x}_{2} & = & [-m_{1}/m+ (\eta \delta m/2m)(v^2-m/r) 
 +(2PN)
+\cdots]\mathbf{x}\,,
\end{eqnarray}
where 
$\eta=\mu/m=m_1m_2/m^2$ is the reduced mass parameter ($0 < \eta \le 1/4$),
and $\delta m = m_1 -m_2$.
The result is a set of equations of motion in terms of relative coordinates:
\begin{equation}\label{eom}
\mathbf{a}=\frac{d^{2}\mathbf{x}}{d t^{2}}
=\frac{m}{r^2}\left [(-1+A)\mathbf{n}+B\mathbf{v}\right ]\,,
\end{equation}
where $A$ and $B$ represent post-Newtonian terms. 
To date,  the two-body equations of motion have been computed up 
to and including 3.5PN order. 
In an appropriate harmonic gauge, 
writing $A=A_{1}+A_{2}+\cdots$ and $B=B_{1}+B_{2}+\cdots$, 
the expressions for $A$ and $B$ read \cite{blanchetiyer03}:
\begin{subequations}
\begin{eqnarray}
A_{1} & = & 2(2+\eta)\frac{m}{r}-(1+3\eta)v^{2}+
\frac{3}{2}\eta{\dot{r}}^{2}\,,\label{A1}\\
A_{2} & = & 
-\frac{3}{4}(12+29\eta){\left (\frac{m}{r}\right )}^{2}
-\eta (3-4\eta)v^{4}
-\frac{15}{8}\eta(1-3\eta){\dot{r}}^{4}
+\frac{1}{2}\eta (13-4\eta)\frac{m}{r}v^{2}
\nonumber\\
& &
+(2+25\eta+2\eta^{2})\frac{m}{r}{\dot{r}}^{2}
+\frac{3}{2}\eta (3-4\eta)v^{2}{\dot{r}}^{2} \,,
\label{A2}\\
A_{5/2} & = & 
\frac{8}{5}\eta\frac{m}{r}\dot{r}\left (\frac{17}{3}\frac{m}{r}+3v^{2} \right )
\,,\label{A52}\\
A_{3} & = & 
\left [16+\left(\frac{1399}{12}
-\frac{41}{16}\pi^{2} \right )\eta
+\frac{71}{2}\eta^{2}\right ]{\left (\frac{m}{r}\right )}^{3}
+\eta\left [\frac{20827}{840}+\frac{123}{64}\pi^{2}-\eta^{2}\right ]
{\left (\frac{m}{r}\right )}^{2}v^{2}
\nonumber\\
& & 
-\left [1+\left (\frac{22717}{168}+\frac{615}{64}\pi^{2}\right )\eta
+\frac{11}{8}\eta^{2}-7\eta^{3}\right ]
{\left (\frac{m}{r}\right )}^{2}{\dot{r}}^{2}
\nonumber\\
& & 
-\frac{1}{4}\eta (11-49\eta+52\eta^{2})v^{6}
+\frac{35}{16}\eta (1-5\eta+5\eta^2 ){\dot{r}}^{6}
- \frac{1}{4}\eta\left (75+32\eta-40\eta^{2}\right )\frac{m}{r}v^{4}
\nonumber\\
& & 
- \frac{1}{2}\eta\left (158-69\eta-60\eta^{2}\right )\frac{m}{r}{\dot{r}}^{4}
+\eta\left (121-16\eta-20\eta^{2}\right )\frac{m}{r}v^{2}{\dot{r}}^{2}
\nonumber\\
& & 
+ \frac{3}{8}\eta\left (20-79\eta+60\eta^{2}\right )v^{4}{\dot{r}}^{2}
-\frac{15}{8}\eta\left (4-18\eta+17\eta^{2}\right )v^{2}{\dot{r}}^{4}
\,,\label{A3}\\
A_{7/2} & = & 
-\frac{8}{5}\eta\frac{m}{r}\dot{r}
\left [\frac{23}{14}(43+14\eta){\left (\frac{m}{r}\right )}^{2}
+\frac{3}{28}(61+70\eta)v^{4}
+70{\dot{r}}^{4}
\right. 
\nonumber\\
& & 
\left. 
+\frac{1}{42}(519-1267\eta)\frac{m}{r}v^{2}
+\frac{1}{4}(147+188\eta)\frac{m}{r}{\dot{r}}^{2}
-\frac{15}{4}(19+2\eta)v^{2}{\dot{r}}^{2}
\right ]\,,
\label{A72}
\end{eqnarray}
\label{Acoeffs}
\end{subequations}
\begin{subequations}
\begin{eqnarray}
B_{1} & = & 2(2-\eta)\dot{r} \,, \label{B1}\\
B_{2} & = & -\frac{1}{2}\dot{r}\left[
(4+41\eta+8\eta^{2})\frac{m}{r}
-\eta(15+4\eta)v^{2}
+3\eta(3+2\eta){\dot{r}}^{2}\right ]\,,
\label{B2}\\
B_{5/2} & = & -\frac{8}{5}\eta\frac{m}{r}\left (3\frac{m}{r}+v^{2}\right )\,,
\label{B52}\\
B_{3} & = & 
\dot{r} \left \{ 
\left [4+\left (\frac{5849}{840}+\frac{123}{32}\pi^{2}\right )\eta
-25\eta^{2}-8\eta^{3}\right ]{\left (\frac{m}{r}\right )}^{2}
+\frac{1}{8} \eta\left(65-152\eta-48\eta^{2}\right )v^{4}
\right .
\nonumber\\
& & 
\left .
+\frac{15}{8}\eta\left (3-8\eta-2\eta^{2}\right ){\dot{r}}^{4}
+\eta\left (15+27\eta+10\eta^{2}\right )\frac{m}{r}v^{2}
\right .
\nonumber\\
& & 
\left .
-\frac{1}{6}\eta\left (329+177\eta+108\eta^{2}\right )\frac{m}{r}\dot{r}^{2}
-\frac{3}{4}\eta\left (16-37\eta-16\eta^{2}\right )v^{2}\dot{r}^{2} \right
\}\,,
\label{B3}\\
B_{7/2}&=&
\frac{8}{5}\eta\frac{m}{r}
\left [
\frac{1}{42}(1325+546\eta){\left (\frac{m}{r}\right )}^{2}
+\frac{1}{28}(313+42\eta)v^{4}
+75\dot{r}^{4}
\right. 
\nonumber\\
& & 
\left. 
-\frac{1}{42}(205+777\eta)\frac{m}{r}v^{2}
+\frac{1}{12}(205+424\eta)\frac{m}{r}\dot{r}^{2}
-\frac{3}{4}(113+2\eta)v^{2}\dot{r}^{2}
\right ]\,.
\label{B72}
\end{eqnarray}
\label{Bcoeffs}
\end{subequations}

At 3PN order, the computation implemented by Blanchet {\it et al.} 
\cite{bf00,bf01} produced logarithmic terms, proportional to 
$\ln(r/{r'}_{1})$ and $\ln(r/{r'}_{2})$, where ${r'}_{1}$ and ${r'}_{2}$ 
are constants related to a scale of radius for each body. 
In obtaining Eqs. (\ref{Acoeffs}) and (\ref{Bcoeffs}), we removed 
these logarithms using a 3PN coordinate
transformation $x_{\mu}\rightarrow x_{\mu}+\delta x_{\mu}$, with 
\cite{bf01}:
\begin{equation}
\delta x_{\mu}=-\frac{22}{3}m_{1}m_{2}\partial_{\mu}
\left [\frac{m_{1}}{y_{2}}\ln\left(\frac{r}{{r'}_{1}}\right )
+\frac{m_{2}}{y_{1}}\ln\left(\frac{r}{{r'}_{2}}\right )\right ],
\label{gaugelog}
\end{equation}
where $y_{a} = |{\bf x} - {\bf x}_a|$ denotes the coordinate separation 
between the considered point and the body $a$.  We note 
that we have $\Box \delta x_{\mu}=0$, except at the location of the two 
bodies. This ensures that the harmonic condition is still respected 
in the new gauge to the required order.   In addition, the parameter
$\lambda$, which was initially undetermined in
\cite{bf00,bf01,blanchetiyer03} has now been fixed to be $\lambda =
-1987/3080$ by different techniques \cite{djs01lett,itohfuta03,bdgef03}; 
that value has
been incorporated into all equations.

In the absence of the 2.5PN and 3.5PN terms, these equations of motion admit
conserved total energy $E$ and total angular momentum $\mathbf{J}$. 
Writing $E=E_{0}+E_{1}+E_{2}+E_{3}$ 
and $\mathbf{J}=\mathbf{J}_{0}+\mathbf{J}_{1}+\mathbf{J}_{2}+\mathbf{J}_{3}$, 
we have:
\begin{subequations}
\begin{eqnarray}
E_{0}/\mu & = & \frac{1}{2}v^{2}-\frac{m}{r},
\label{E0}\\
E_{1}/\mu & = & 
\frac{1}{2}{\left (\frac{m}{r}\right )}^{2}
+\frac{3}{8}(1-3\eta)v^{4}
+\frac{1}{2}(3+\eta)v^{2}\frac{m}{r}
+\frac{1}{2}\eta\frac{m}{r}\dot{r}^{2} \,,
\label{E1} \\
E_{2}/\mu & = & 
-\frac{1}{4}(2+15\eta){\left (\frac{m}{r}\right )}^{3}
+\frac{5}{16}(1-7\eta+13\eta^{2})v^{6}
+\frac{1}{8}(14-55\eta+4\eta^{2}){\left(\frac{m}{r}\right)}^{2}v^{2}
\nonumber\\
 & & 
+\frac{1}{8}(4+69\eta+12\eta^{2}){\left (\frac{m}{r}\right )}^{2}\dot{r}^{2}
+\frac{1}{8}(21-23\eta-27\eta^{2})\frac{m}{r}v^{4}
\nonumber\\
 & & 
+\frac{1}{4}\eta (1-15\eta)\frac{m}{r}v^{2}\dot{r}^{2}
-\frac{3}{8}\eta (1-3\eta)\frac{m}{r}\dot{r}^{4} \,,
\label{E2}
\\
E_{3}/\mu&=&
\left [\frac{3}{8}+ \frac{18469}{840}\eta\right ]{\left (\frac{m}{r}\right )}^{4}
 +\left [\frac{5}{4}-\left (\frac{6747}{280}-\frac{41}{64}\pi^{2}\right )\eta-\frac{21}{4}\eta^{2}+\frac{1}{2}\eta^{3}\right ]{\left (\frac{m}{r}\right )}^{3}v^{2}
\nonumber\\
 & & 
 +\left [\frac{3}{2}+\left (\frac{2321}{280}-\frac{123}{64}\pi^{2}\right )\eta+\frac{51}{4}\eta^{2} +\frac{7}{2}\eta^{3}\right ]{\left (\frac{m}{r}\right )}^{3}\dot{r}^{2}
 \nonumber\\
 & &
+\frac{1}{128}\left (35-413\eta+1666\eta^{2}-2261\eta^{3}\right )v^{8}
 +\frac{1}{16}(135-194\eta+406\eta^{2}-108\eta^{3}){\left (\frac{m}{r}\right )}^{2}v^{4}
 \nonumber\\
 & & 
 +\frac{1}{16}(12+248\eta-815\eta^{2}-324\eta^{3}){\left (\frac{m}{r}\right )}^{2}v^{2}\dot{r}^{2}
 -\frac{1}{48}\eta(731-492\eta-288\eta^{2}){\left (\frac{m}{r}\right )}^{2}\dot{r}^{4}
 \nonumber\\
 & & 
 +\frac{1}{16}(55-215\eta+116\eta^{2}+325\eta^{3})\frac{m}{r}v^{6}
 +\frac{1}{16}\eta(5-25\eta+25\eta^{2})\frac{m}{r}\dot{r}^{6}
 \nonumber\\
 & & 
 -\frac{1}{16}\eta(21+75\eta-375\eta^{2})\frac{m}{r}v^{4}\dot{r}^{2}
 -\frac{1}{16}\eta(9-84\eta+165\eta^{2})\frac{m}{r}v^{2}\dot{r}^{4},
 \label{E3}
 \end{eqnarray}
 \label{Energy3pn}
 \end{subequations}
 \begin{subequations}
 \begin{eqnarray}
\mathbf{J}_{0}/\mu &=& (\mathbf{x}\times\mathbf{v}) \,,
\label{J0} \\
\mathbf{J}_{1}/\mu &=& (\mathbf{x}\times\mathbf{v})\left [(3+\eta)\frac{m}{r}+\frac{1}{2}(1-3\eta)v^{2}\right ] \,,
\label{J1} \\
\mathbf{J}_{2}/\mu &=& (\mathbf{x}\times\mathbf{v})\left [
\frac{1}{4}(14-41\eta+4\eta^{2}){\left (\frac{m}{r}\right )}^{2}
+\frac{3}{8}(1-7\eta+13\eta^{2})v^{4}
\right.
\nonumber\\
 & &
  \left.
 +\frac{1}{2}(7-10\eta-9\eta^{2})\frac{m}{r}v^{2}
 -\frac{1}{2}\eta(2+5\eta)\frac{m}{r}\dot{r}^{2}
 \right ],
 \label{J2} \\
\mathbf{J}_{3}/\mu &=& (\mathbf{x}\times\mathbf{v})\left \{
\left [ \frac{5}{2}-\left (\frac{5199}{280}-\frac{41}{32}\pi^{2}\right )\eta-7\eta^{2}+\eta^{3}\right ]{\left (\frac{m}{r}\right )}^{3}
\right. 
\nonumber\\
 & & 
 \left. 
 +\frac{1}{16}(5-59\eta+238\eta^{2}-323\eta^{3} )v^{6}
+\frac{1}{12}(135-322\eta+315\eta^{2}-108\eta^{3}){\left (\frac{m}{r}\right )}^{2}v^{2}
\right.
\nonumber\\
 & &
  \left.
+\frac{1}{24} (12-287\eta-951\eta^{2}-324\eta^{3}){\left (\frac{m}{r}\right )}^{2}\dot{r}^{2}
+\frac{1}{8}(33-142\eta+106\eta^{2}+195\eta^{3})\frac{m}{r}v^{4}
\right. 
\nonumber\\
 & & 
 \left.
 -\frac{1}{4} \eta (12-7\eta-75\eta^2)\frac{m}{r}v^{2}\dot{r}^{2}
+\frac{3}{8}\eta(2-2\eta-11\eta^{2})\frac{m}{r}\dot{r}^{4}\right \} \,.
\label{J3}
\end{eqnarray}
\label{angmomentum3pn}
\end{subequations}

\subsection{Solution of the 3PN equations of motion}
\label{osculate}

In order to solve these equations, we shall initially adopt the method of 
osculating orbital elements, which is well-adapted to the perturbed two-body 
Kepler problem. The osculating orbit elements are defined by the 
Keplerian orbit that is tangent to the actual trajectory at a 
particular moment of time. In the Newtonian case, the osculating elements 
are constants of the motion; in a perturbed Newtonian problem, they 
change smoothly with time (see \cite{lincwill} for more details about the method of osculating elements applied to the post-Newtonian problem).

From the equations of motion we can easily deduce that the 
trajectory is planar, which allows us to reduce the number of variables 
from six to four. If we assume that the plane of the motion is 
perpendicular to $\mathbf{\hat z}$ ($x,y,z$ being a standard cartesian 
coordinate system), our new set of variables $(\alpha,\beta,p,\phi)$ 
is related to the old set
$(x,y,v_{x},v_{y})$ by the definitions (some of which are redundant):
\begin{eqnarray}
x & \equiv & r\cos\phi\,,
\nonumber \\
y & \equiv & r\sin\phi\,,
\nonumber \\
v_{x}&\equiv&-(m/p)^{1/2}(\beta+\sin\phi)\,,
\nonumber \\
v_{y}&\equiv&(m/p)^{1/2}(\alpha+\cos\phi)\,,
\nonumber \\
r & = & p(1+\alpha\cos\phi+\beta\sin\phi)^{-1} \,,
\nonumber \\
r^{2}\dot{\phi} &\equiv& (mp)^{1/2} \,.
\label{xyvxvy}
\end{eqnarray}
Reciprocally, we can deduce the osculating elements from the orbital 
variables by using the following relations:
\begin{eqnarray}
\phi & = & \arctan\left (\frac{y}{x}\right )\,,
\nonumber \\
\alpha & = & \frac{\sigma}{m}v_{y}-\frac{x}{r}\,,
\nonumber \\
\beta & = & -\frac{\sigma}{m}v_{x}-\frac{y}{r}\,,
\nonumber \\
p & = & \sigma^2/m\,,
\nonumber \\
\sigma & = &(\mathbf{x}\times\mathbf{v})\cdot \mathbf{\hat z} \,.
\label{alphabetadefs}
\end{eqnarray}
One additional expression will be useful:
\begin{equation}\label{dotr}
\dot{r}=(m/p)^{1/2}(\alpha\sin\phi-\beta\cos\phi) \,.
\end{equation}
We note that the 
vector $(\alpha,\beta)$ has as its norm the ordinary Keplerian osculating
eccentricity $e$ and as its phase angle
the direction $\omega$ of the Keplerian osculating
periastron, so that we have $\alpha=e\cos\omega$ and $\beta=e\sin\omega$.

In what follows, we will use the parameter $u=m/p$ rather than $p$. 
Note that $u$ is of order $\epsilon \sim m/r$. 
In the Newtonian case, $u$, $\alpha$ and $\beta$ are constants of the motion; 
in the post-Newtonian problem, these parameters vary according 
to the following ``Lagrange 
planetary equations'' (so-called from their extensive use
in solar-system studies):
\begin{eqnarray}
\frac{d u}{d\phi} & = & -2u^{3/2}B,
\nonumber \\
\frac{d \alpha}{d\phi} & = & A\sin\phi+2u^{1/2}B(\alpha+\cos\phi ),
\nonumber \\
\frac{d \beta}{d\phi} & = & -A\cos\phi+2u^{1/2}B(\beta+\sin\phi ),
\label{evoluab}
\end{eqnarray}
where we have used Eqs. (\ref{eom}), (\ref{xyvxvy}) and (\ref{dotr}). 
When the definitions of $\mathbf{x}$ and $\mathbf{v}$ 
[Eqs. (\ref{xyvxvy})] are substituted into the PN expressions for
$A$ and $B$ [Eqs. (\ref{Acoeffs}) and  (\ref{Bcoeffs})], we get a set of 
coupled first-order differential equations in 
the variables $\alpha(\phi)$, $\beta(\phi)$ and $u(\phi)$.

The planetary equations derived from Eqs. (\ref{evoluab}) 
are too long to be reproduced here (they can be found through 2.5PN 
order in \cite{lincwill}). However we can schematically write 
them in the general form:
\begin{eqnarray}
\frac{d u}{d\phi}&=&u^2\mathcal{D}u_{1}(\alpha,\beta,{\phi})
+u^3\mathcal{D}u_{2}(\alpha,\beta,{\phi})
+u^{5/2}\mathcal{D}u_{5/2}(\alpha,\beta,{\phi})+\cdots \,,
\nonumber \\
\frac{d \alpha}{d\phi}&=&u\mathcal{D}\alpha_{1}(\alpha,\beta,{\phi})
+u^{2}\mathcal{D}\alpha_{2}(\alpha,\beta,{\phi})
+u^{5/2}\mathcal{D}\alpha_{5/2}(\alpha,\beta,{\phi})
+\cdots \,,
\nonumber \\
\frac{d \beta}{d\phi}&=&u\mathcal{D}\beta_{1}(\alpha,\beta,{\phi})
+u^{2}\mathcal{D}\beta_{2}(\alpha,\beta,{\phi})
+u^{5/2}\mathcal{D}\beta_{5/2}(\alpha,\beta,{\phi})
+\cdots  \,,
\label{evoluab1}
\end{eqnarray}
where $\mathcal{D}u_{n}$, $\mathcal{D}\alpha_{n}$ and 
$\mathcal{D}\beta_{n}$ ($n\in \{1,2,5/2,3,7/2\}$) are  
polynomials in $\alpha$ and $\beta$ and simple trigonometric 
functions of $\phi$. 
We quote, for illustration, the first post-Newtonian expressions for
these polynomials:
\begin{eqnarray}
\mathcal{D}u_{1} &=& 4(2-\eta)(\beta \cos \phi - \alpha \sin \phi )
\,,
\nonumber \\
\mathcal{D}\alpha_{1} &=& - 3\beta + (3-\eta) \sin \phi 
+(5-4\eta) (\alpha \sin 2\phi - \beta \cos 2\phi)
\nonumber \\
&&
+\frac{1}{8}[(56-47\eta)\alpha^2-(8+21\eta)\beta^2]\sin \phi
-\frac{1}{4}(32-13\eta)\alpha\beta\cos\phi
\nonumber \\
&&
+\frac{3}{8}\eta(\beta^2-\alpha^2)\sin 3\phi
+\frac{3}{4}\eta\alpha\beta\cos 3\phi \,,
\nonumber \\
\mathcal{D}\beta_{1} &=& 3\alpha -(3-\eta) \cos \phi 
-(5-4\eta) (\alpha \cos 2\phi + \beta \sin 2\phi)
\nonumber \\
&&
-\frac{1}{8}[(56-47\eta)\beta^2-(8+21\eta)\alpha^2]\cos \phi
+\frac{1}{4}(32-13\eta)\alpha\beta\sin\phi
\nonumber \\
&&
+\frac{3}{8}\eta(\alpha^2-\beta^2)\cos 3\phi
+\frac{3}{4}\eta\alpha\beta\sin 3\phi \,.
\label{firstorderpolynom}
\end{eqnarray}

We want to solve these equations iteratively. 
At zeroth (Newtonian) order $u$, $\alpha$ and $\beta$ are constants of the
motion $\tilde u$, $\tilde\alpha$ and $\tilde\beta$, 
and can be related to the initial state of the orbit.  
Post-Newtonian effects cause them to vary slowly over a post-Newtonian
timescale or a radiation-reaction timescale, related to the orbital phase
$\phi$ by $\phi/\epsilon$ and $\phi/\epsilon^{5/2}$, respectively.
Superimposed upon this will be variations on an orbital timescale.  To take
these two effects into account, we use a two-scale approach \cite{bender}.
We define a variable $\theta = \epsilon \phi$, and we assume that 
the osculating elements 
can be written as functions of $\theta$ and $\phi$ in the
generic form $u = u({\tilde u}(\theta), {\tilde\alpha}(\theta), {\tilde
\beta}(\theta), \phi)$, with $\theta$ and $\phi$ now treated as independent
variables.  We then expand 
the elements in powers of $\epsilon$:
\begin{eqnarray}
u&=&\epsilon \tilde{u}
+\epsilon^2 u_{1}(\tilde{\alpha},\tilde{\beta},\tilde{u},\phi)
+\epsilon^{3} u_{2}(\tilde{\alpha},\tilde{\beta},\tilde{u},\phi)
+\cdots \,,
\nonumber \\
\beta&=&\tilde{\beta}
+\epsilon\beta_{1}(\tilde{\alpha},\tilde{\beta},\tilde{u},\phi)
+\epsilon^{2}\beta_{2}(\tilde{\alpha},\tilde{\beta},\tilde{u},\phi)
+\cdots\,,
\nonumber \\
\alpha&=&\tilde{\alpha}
+\epsilon\alpha_{1}(\tilde{\alpha},\tilde{\beta},\tilde{u},\phi)
+\epsilon^{2}\alpha_{2}(\tilde{\alpha},\tilde{\beta},\tilde{u},\phi)
+\cdots\,.
\label{expansion}
\end{eqnarray}
Notice that, by its very nature, $u$ begins at order $\epsilon$.
We write the derivative with respect to $\phi$ in the form
\begin{eqnarray}
\frac{d}{d \phi} &=&
\frac{\partial}{\partial {\phi}}
+ \epsilon \frac{\partial}{\partial {\theta}}
 \nonumber \\
 &=& 
\frac{\partial}{\partial {\phi}}+
\epsilon \left (
\frac{d \tilde{\alpha}}{d \theta}\frac{\partial}{\partial \tilde{\alpha}}
+\frac{d \tilde{\beta}}{d \theta}\frac{\partial}{\partial \tilde{\beta}}
+\frac{d \tilde{u}}{d \theta}\frac{\partial}{\partial \tilde{u}}
\right ) \,.
\end{eqnarray}

We also expand the derivatives with respect to 
$\theta$ in powers of $\epsilon$:
\begin{eqnarray}
\frac{d \tilde{u}}{d \theta}
&=& d\tilde{u}_{1}(\tilde{\alpha},\tilde{\beta},\tilde{u})
+\epsilon d\tilde{u}_{2}(\tilde{\alpha},\tilde{\beta},\tilde{u})+\cdots
\,,
\nonumber \\
\frac{d \tilde{\beta}}{d \theta}
&=&d\tilde{\beta}_{1}(\tilde{\alpha},\tilde{\beta},\tilde{u})
+\epsilon d\tilde{\beta}_{2}(\tilde{\alpha},\tilde{\beta},\tilde{u})
+\cdots
\,,
\nonumber \\
\frac{d \tilde{\alpha}}{d \theta}
&=&d\tilde{\alpha}_{1}(\tilde{\alpha},\tilde{\beta},\tilde{u})
+\epsilon d\tilde{\alpha}_{2}(\tilde{\alpha},\tilde{\beta},\tilde{u})+\cdots
\,.
\end{eqnarray}
Now we have reduced our study to the search for $\alpha_{i}$, $\beta_{i}$, 
$u_{i}$ on the one hand, which will give the dependence on 
$\tilde{\alpha}$, $\tilde{\beta}$, $\tilde{u}$, and $\phi$, 
and $d\tilde{\alpha}_{i}$, $d\tilde{\beta}_{i}$, $d\tilde{u}_{i}$ 
on the other hand, which will give differential equations allowing solution
for the $\theta$-dependence, or long-term variation of the parameters. 
Note that this is not the only way to decompose the problem, but is a 
natural way, given the split into orbital and secular evolution of the
variables.

We now define the average and the average-free part of a 
function $f({\phi})$ by:
\begin{eqnarray}
\langle f\rangle&=&
\frac{1}{2\pi}\int_{0}^{2\pi}f({\phi})d {\phi},
\nonumber \\
\mathcal{AF}(f)({\phi})&=&f({\phi})-\langle f\rangle,
\end{eqnarray}
where the ``independent'' variable $\theta$ is held fixed.  (An equivalent procedure
would be to convert all functions of $\phi$ into $2\pi$-periodic functions
and constants.)
We rewrite Eqs. (\ref{evoluab}) with our new variables, 
and we collect terms of common powers of $\epsilon$. 
At first order in $\epsilon$ we get 
\begin{eqnarray}
d\tilde{u}_{1}+\frac{\partial u_{1}}{\partial {\phi}}
&=&
\tilde{u}^2 \mathcal{D}u_1(\tilde{\alpha},\tilde{\beta},\phi) \,,
\nonumber \\
d\tilde{\alpha}_{1}+\frac{\partial \alpha_{1}}{\partial {\phi}}
&=& \tilde{u}\mathcal{D}\alpha_{1}(\tilde{\alpha},\tilde{\beta},{\phi})
\,,
\nonumber \\
d\tilde{\beta}_{1}+\frac{\partial \beta_{1}}{\partial {\phi}}
&=& \tilde{u}\mathcal{D}\beta_{1}(\tilde{\alpha},\tilde{\beta},{\phi})
\,.
\label{firstorder}
\end{eqnarray}
where the expressions on the right-hand-side are given by Eqs.
(\ref{firstorderpolynom}), with $\tilde{\alpha}$ replacing $\alpha$, and so
on.
Reading off the average parts of Eqs. (\ref{firstorder}), 
we find $d\tilde{u}_{1} =0$, 
$d\tilde{\alpha}_{1} = -3\tilde{u}\tilde{\beta}$, and 
$d\tilde{\beta}_{1} = 3\tilde{u}\tilde{\alpha}$.  Defining 
$\tilde{\alpha} \equiv \tilde{e} \cos \tilde{\omega}$ and  
$\tilde{\beta} \equiv \tilde{e} \sin \tilde{\omega}$ 
we find, to first PN order that
\begin{eqnarray}
d\tilde{u}/d\theta &=& 0  \,, 
\nonumber \\
d\tilde{e}/d\theta &=& (\tilde{\alpha} d\tilde{\alpha}/d\theta
 +\tilde{\beta} d\tilde{\beta}/d\theta )/\tilde{e} = 0 \,,
 \nonumber \\
d\tilde{\omega}/d\theta &=& (\tilde{\alpha} d\tilde{\beta}/d\theta
-\tilde{\beta} d\tilde{\alpha}/d\theta)/\tilde{e}^2 
= 3\tilde{u}  \,.
\end{eqnarray}
These results express the well-known fact that the orbital eccentricity and
semi-latus rectum do not evolve secularly to 1PN order; in fact, this holds
true at 2PN and 3PN order; they only evolve secularly as a result of
radiation reaction.  The angle of pericenter $\tilde{\omega}$
evolves secularly at 1PN order via the standard advance; there are
also 2PN and 3PN contributions, but no radiation-reaction contributions to
the advance of $\tilde{\omega}$, through 3.5PN order.

Then, integrating the average-free parts of Eqs. (\ref{firstorder}), 
we obtain, for
example,
\begin{equation}
\alpha_{1}=\mathcal{AF}\left (\int \mathcal{AF}(\mathcal{D}\alpha_{1})({\phi})
d{\phi}\right ).
\end{equation}
The role of the second $\mathcal{AF}$ is to get rid 
of the constant of integration.
The same method yields similar results for $\beta_1$ and $u_1$.

At second order in $\epsilon$, we obtain equations of the form
\begin{eqnarray}
d\tilde{\alpha}_{2}+\frac{\partial \alpha_{2}}{\partial {\phi}}
&=&\tilde{u}^2 \mathcal{D}\alpha_{2}
+\tilde{u} \left (
\frac{\partial \mathcal{D}\alpha_{1}}{\partial \alpha}\alpha_{1}
+\frac{\partial \mathcal{D}\alpha_{1}}{\partial \beta}\beta_{1}
\right )
\nonumber \\
&&+ u_1 \mathcal{D}\alpha_{1}
-\frac{\partial \alpha_{1}}{\partial \tilde{\alpha}}d\tilde{\alpha}_{1}
-\frac{\partial \alpha_{1}}{\partial \tilde{\beta}}d\tilde{\beta}_{1}
-\frac{\partial \alpha_{1}}{\partial \tilde{u}}d\tilde{u}_{1}
\nonumber \\
&\equiv & f_{2}(\tilde{\alpha},\tilde{\beta},\tilde{u},{\phi}),
\end{eqnarray}
where $\alpha_{1}$, $\beta_{1}$, $u_{1}$, $d\tilde{\alpha}_{1}$, $d\tilde{\beta}_{1}$ and
$d\tilde{u}_{1}$ are known from the first order solution. 
For the same reasons as previously we have:
\begin{eqnarray}
d\tilde{\alpha}_{2}&=&\langle f_{2} \rangle \,,
\nonumber \\
\alpha_{2}&=&\mathcal{AF}\left (\int \mathcal{AF}(f_{2})({\phi})d{\phi}\right ).
\end{eqnarray}
Using this procedure systematically up to 3.5PN order, 
we completely determine 
$\alpha(\tilde{\alpha},\tilde{\beta},\tilde{u},{\phi})$, 
$\beta(\tilde{\alpha},\tilde{\beta},\tilde{u},{\phi})$ 
and $u(\tilde{\alpha},\tilde{\beta},\tilde{u},{\phi})$, 
as well as 
$\frac{d \tilde{\alpha}}{d \phi}(\tilde{\alpha},\tilde{\beta},\tilde{u})$, 
$\frac{d \tilde{\beta}}{d \phi}(\tilde{\alpha},\tilde{\beta},\tilde{u})$ 
and $\frac{d \tilde{u}}{d \phi}(\tilde{\alpha},\tilde{\beta},\tilde{u})$. 
From this and Eqs. (\ref{xyvxvy}) 
we can deduce the explicit expressions for
$\mathbf{x}$, $\mathbf{v}$, $r$, \emph{etc.}

To 3.5PN order, the secular evolution of $\tilde{u}$ and $\tilde{e}$ is
governed by radiation reaction, and is given by the coupled equations 
(we now set $\epsilon = 1$)
\begin{eqnarray}
\frac{d \tilde{u}}{d \phi}
&=&\frac{8}{5}\eta \biggl \{ (8+7\tilde{e}^{2})\tilde{u}^{7/2}
-\left [ \left ( \frac{2759}{42}+6\eta \right)
\right.
\nonumber\\
&&
\left.
-\left ( \frac{379}{21}+\frac{127}{6}\eta \right )\tilde{e}^{2}
-\left (\frac{1483}{336}+\frac{79}{6}\eta \right )\tilde{e}^{4} 
\right ]\tilde{u}^{9/2} \biggr \} \,,
\nonumber
\\
\frac{d \tilde{e}}{d \phi}
&=&-\frac{1}{15}\eta\tilde{e} \biggl \{ (304+121\tilde{e}^{2} )\tilde{u}^{5/2}
- \left [\left(\frac{18049}{7}+636\eta \right )
\right.
\nonumber \\
&&
\left.
-\left (\frac{4346}{7}+\frac{1829}{2}\eta \right ) \tilde{e}^{2}
-\left (\frac{2251}{56}+269\eta\right ) \tilde{e}^{4} \right ]
\tilde{u}^{7/2} 
\biggr \}\,.
\label{evoleut}
\end{eqnarray}

We note that the eccentricity decreases as the orbit shrinks. 
The periastron advance is driven by the conservative part of the equations:
\begin{eqnarray}
\label{evolomega}
\frac{d \tilde{\omega}}{d \phi}
&=&3\tilde{u}-\frac{3}{4}\left [10+4\eta-(1+10\eta)\tilde{e}^{2}\right ]
\tilde{u}^{2}
+\left \{\frac{87}{2}-\left (\frac{157}{4}-\frac{123}{32}\pi^{2}\right )\eta
-3\eta^{2}
\right.
\nonumber \\
&&
\left.
-\left [45-\left(23+\frac{123}{128}\pi^{2}\right )\eta
+\frac{93}{2}\eta^{2}\right ]\tilde{e}^{2}
+\frac{3}{8}\eta(12-25\eta)\tilde{e}^{4}\right \}\tilde{u}^{3}
\,.
\end{eqnarray}

\subsection{Energy and angular momentum in terms of new orbit elements}
\label{conserved}

We now wish to convert from the osculating orbit elements $\tilde{u}$ 
and $\tilde{e}$
to our alternative quantities defined in Eqs. (\ref{ezeta}) (cf. Sec
\ref{turningpoint}).  
Using the formula
\begin{equation}
m\Omega \equiv m^{3/2}p^{1/2}/r^2 = {u}^{3/2} (1+{\alpha} \cos \phi
+{\beta} \sin \phi )^2 \,,
\end{equation}
we can easily show that the maxima and minima of $\Omega$ occur at $\phi =
\tilde{\omega}$ (pericenter) and $\phi =
\tilde{\omega} + \pi$, (apocenter) respectively.  We then
express $\Omega_{p}$ and $\Omega_{a}$, and thence our new
orbit elements $e$ and $\zeta$ as functions 
of $\tilde{e}$ and $\tilde{u}$.  To 2PN order, the relationships are given by
\begin{subequations}
\begin{eqnarray}
{e} &=& \tilde{e} \left \{ 1 + \left [\frac{1}{2} (13-4\eta) + (1-3\eta)
\tilde{e}^2 \right ] \tilde{u}
+ \left [\frac{1}{4} (52 -129\eta) 
\right .
\right .
\nonumber \\
&& 
\left .
\left .
+ \frac{1}{16} (157-337\eta+116\eta^2) 
\tilde{e}^2 +  \frac{1}{4} (4-19\eta+48\eta^2)\tilde{e}^4 \right ]
\tilde{u}^2 \right \} \,,
\label{e3PN}
\\
\zeta&=& \tilde{u} \left \{ 1 - \frac{4}{3} \left [ (3-\eta) +
 (1-3\eta) \tilde{e}^2 \right ] \tilde{u}
+ \left [ \frac{1}{9} (198+39\eta+26\eta^2)
\right .
\right .
\nonumber \\
&& 
\left .
\left .
- \frac{1}{36} (1092-977\eta+276\eta^2) \tilde{e}^2 
+ \frac{1}{9} (2-27\eta-18\eta^2) \tilde{e}^4 \right ]
\tilde{u}^2 \right \} \,.
\label{zeta3PN}
\end{eqnarray}
\label{ezeta3PN}
\end{subequations}
Notice that a circular orbit corresponds to $\tilde{e}={e}=0$.

We invert these relations and substitute 
the expressions for $\tilde{e}(e,\zeta)$ and 
$\tilde{u}(e,\zeta)$ into the solution of the equations of motion. 
The results for $m/r$ and $r^{2}\dot{\phi}$ to 3PN 
order are too long to be reproduced here. However, in order to give 
an idea of what they look like, we quote them to 2PN order, expressed in
terms of our new orbit elements.  
\begin{eqnarray}
\frac{{m}}{r} & = & \biggl \{ 1+e\cos \phi^\prime \biggr \} \zeta 
 + \left \{ 1-\frac{1}{3}\eta+ \frac{7}{12} (4-3\eta){e}^2
 \right .
\nonumber\\ 
& & 
\left .
 +\frac{1}{3} \left [ (9-4\eta) +(1-3\eta){e}^{2} 
 \right ] {e}\cos\phi^\prime
-\frac{\eta}{4}e^{2}\cos (2\phi^\prime)\right \}\zeta^{2} 
\nonumber\\
& &
 + \left \{ 1-\frac{65}{12}\eta + \frac{1}{24}(356-319\eta+48\eta^2)e^2
  +\frac{1}{192}(256-265\eta+459\eta^2)e^4
\right .
\nonumber\\
& &
\left .
 + \left [ \frac{1}{12}(96-231\eta+8\eta^2) +
 \frac{1}{48}(323-351\eta+180\eta^2)e^2 
 + \frac{\eta}{12}(5+12\eta)e^4 \right ] e \cos \phi^\prime
 \right .  
 \nonumber\\
 & &
 \left .
 - \left [ \frac{1}{24}(60+159\eta-16\eta^2)
 + \frac{\eta}{48}(29-27\eta)e^2 \right ] e^2 \cos (2\phi^\prime)
\right .  
\nonumber\\
& &
\left .
 -\frac{1}{16}(1+19\eta-4\eta^2)e^3 \cos (3\phi^\prime)
 -\frac{\eta}{64}(1-3\eta)e^4\cos (4\phi^\prime)
 \right \} \zeta^3
\,,
\nonumber
\\
r^2\dot{\phi} & = & \frac{m}{\sqrt{\zeta}}\left [
1-\left \{
\frac{2}{3}(3-\eta) + \frac{2}{3}(1-3\eta)e^2 +
(4-2\eta)e\cos \phi^\prime
\right \}\zeta 
\right .
\nonumber\\
& &
\left .
 + \left \{ \frac{1}{6}(6+53\eta+2\eta^2) -
\frac{1}{24}(28+117\eta-12\eta^2)e^2
+\frac{1}{6}(2-17\eta+6\eta^2)e^4
\right .
\right .
\nonumber\\
& &
\left .
\left .
- \left [
\frac{1}{3}(6-53\eta-2\eta^2)-\frac{1}{24}(32-211\eta+54\eta^2)e^2 \right ]
e \cos \phi^\prime
\right .
\right .
\nonumber\\
& &
\left .
\left .
+ \frac{1}{8}(36-13\eta+4\eta^2)e^2 \cos (2\phi^\prime)
-\frac{\eta}{8}(3+2\eta)e^3 \cos (3\phi^\prime)
\right \} \zeta^2
\right ]
\,,
\label{orbitsol}
\end{eqnarray}
where $\phi^\prime \equiv \phi - \omega$.  
The leading term corresponds to the Newtonian solution. 
Note that $\zeta$, $e$ and 
$\omega$ are now our \emph{post-Newtonian} orbital 
elements, and should not be mistaken for the \emph{Newtonian} 
$u$, $e$ and $\omega$ introduced in Sec. \ref{turningpoint}. 

An alternative method for integrating the post-Newtonian equations of motion
was developed by
Wagoner and Will \cite{wagwill}.  In that method, perturbations of the
velocity and angular momentum were defined by the equations 
\begin{eqnarray}
r^2 d\phi/dt &\equiv& |\mathbf{x} \times \mathbf{v}| \equiv (mp)^{1/2} (1+ \delta h) \,,\nonumber \\
\mathbf{v} &\equiv& (m/p)^{1/2} [-\sin (\phi-\omega)\mathbf{e_x} 
+ (\hat{e}+\cos (\phi-\omega))\mathbf{e_y}
+ \delta\mathbf{v} ) \,,
\end{eqnarray}
where $p$, $\hat{e}$ and $\omega$ are constants.
Taking a time derivative of both equations, substituting
the 3PN equations of motion (ignoring radiation reaction terms), converting
to derivatives with respect to $\phi$ and integrating, one obtains
expressions for the perturbed $r^2 d\phi/dt$ and $\mathbf{v} $.  One then
integrates the identity $(d/d\phi)r^{-1} = -(r^2d\phi/dt)^{-1} 
(\mathbf{n} \cdot \mathbf{v} )$ with respect to $\phi$, setting the
constants of integration at each PN order so that the identity 
$d(r\mathbf{n})/dt =
\mathbf{v}$ is reproduced.  
In terms of the bare orbit elements $\hat{e}$, $\hat{u} = m/p$
and $\omega$, the orbit equations look different at each PN order from those
derived above in terms of $\tilde{e}$, $\tilde{u}$ and $\tilde{\omega}$.
But when the solution so derived 
is used to identify $\Omega_a$ and $\Omega_p$ and 
thence to
define new orbit elements from Eqs. (\ref{ezeta}), the resulting orbit
solution in terms of our 
new orbit elements is {\it identical} 
to Eqs. (\ref{orbitsol}), through 3PN order.

The equations describing the evolution
with time of our new orbit elements then
become:
\begin{eqnarray}
\frac{d \zeta}{d \phi}
&=&\frac{8}{5}\eta \biggl \{ (8+7e^{2} )\zeta^{7/2}
-\left [ \frac{743}{42}+22\eta
\right.
\nonumber\\
&&
\left.
-\left (\frac{1186}{63}-\frac{685}{6}\eta \right )e^{2}
-\left (\frac{18001}{1008}-\frac{163}{6}\eta \right )e^{4}\right ]\zeta^{9/2}
\biggr \} \,,
\nonumber \\
\frac{d e}{d \phi}
&=&-\frac{1}{15}\eta e\biggl \{ (304+121{e}^{2}){\zeta}^{5/2}
-\left [  \frac{5505}{7}+\frac{3796}{3} \eta
\right.
\nonumber\\
&&
\left.
-\left (\frac{12499}{21}-\frac{17741}{6}\eta \right ){e}^{2}
-\left (\frac{46289}{168}-437\eta \right )e^{4}
\right ]\tilde{\zeta}^{7/2}
\biggr \} \,,
\nonumber \\
\frac{d \omega}{d \phi}&=&
3\zeta+\left[\frac{1}{2}(9-14\eta)+\frac{1}{4}(19-18\eta )e^2
\right ]\zeta^{2} 
+\left \{\frac{27}{2}-\left (\frac{481}{4}-\frac{123}{32}\pi^{2}\right )\eta
+7\eta^{2}
\right.
\nonumber\\
&&\left.
+\left [\frac{137}{4}-\left (\frac{337}{4}-\frac{123}{128}\pi^{2}
\right )\eta+\frac{53}{2}\eta^{2}\right ]e^{2}
+\frac{1}{8}(20+8\eta+45\eta^{2} )e^{4}\right \}
\zeta^{3} \,.
\label{evolh}
\end{eqnarray}
Now the problem is entirely solved. Equations
(\ref{orbitsol}) pushed to 3PN order, characterize the 
motion, while Eqs. (\ref{evolh}) 
give the pericenter advance and effect of radiation reaction on the 
orbital elements.

We now ignore the effects of radiation reaction, express all the orbital
variables $r$, $\mathbf{x}$, $\mathbf{v}$, $\dot r$ to 3PN order in terms of
our new orbit elements and the angle $\phi$, and substitute
into the expressions (\ref{Energy3pn}) and (\ref{angmomentum3pn}).  As
expected $E$ and $\mathbf{J}$ are constant (independent of $\phi$) through
3PN order.  Defining $\tilde{E}=E/\mu$ and $\tilde{J}=|J|/\mu m$, with
$\mathbf{J}=|J| \mathbf{\hat z}$, we find for a general eccentric orbit:
\begin{subequations}
\begin{eqnarray}
\tilde{E}_{\textrm{Harm}}&=&-\frac{1}{2}(1-e^{2})\zeta
\left \{ 1 -\left [ \frac{3}{4}+\frac{1}{12}\eta-\left (\frac{1}{12}-\frac{1}{4}\eta\right )e^{2}
\right ]\zeta
\right .
\nonumber\\
 &&
 \left .
 -\left [\frac{27}{8}-\frac{19}{8}\eta+\frac{1}{24}\eta^{2}
-\left (\frac{17}{12}+4\eta+\frac{1}{4}\eta^{2}\right )e^{2}
+\left (\frac{1}{24}+\frac{29}{24}\eta-\frac{1}{8}\eta^{2}\right )e^{4}
\right ]\zeta^{2}
\right.
\nonumber\\
&&
\left.
-\left [ \frac{675}{64}-\left (\frac{34445}{576}-\frac{205}{96}\pi^{2} 
\right)\eta
+\frac{155}{96}\eta^{2}+\frac{35}{5184}\eta^{3}
\right.
\right.
\nonumber\\
&&
\left.
\left.
+\left (\frac{7}{64}-\left (\frac{2369}{576}+\frac{41}{96}\pi^{2}\right )\eta
+\frac{11951}{864}\eta^{2}-\frac{25}{576}\eta^{3}\right )e^{2}
\right.
\right.
\nonumber\\
&&
\left.
\left.
-\left (\frac{815}{576}-\frac{7619}{1728} \eta
-\frac{1499}{288}\eta^{2}-\frac{25}{64}\eta^{3}\right )e^{4}
\right.
\right.
\nonumber\\
&&
\left.
\left.
-\left(\frac{35}{5184}-\frac{143}{192}\eta+\frac{57}{32}\eta^{2}
-\frac{5}{64}\eta^{3}\right )e^{6}
\right ]\zeta^{3} \right \} \,,
\label{Eharm}
\\
\tilde{J}_{\textrm{Harm}}&=&\frac{1}{\sqrt{\zeta}}\left 
\{1+\left[ \frac{3}{2}+\frac{1}{6}\eta-\left (
\frac{1}{6}-\frac{1}{2}\eta\right )e^{2}\right ]\zeta
\right.
\nonumber\\
&&
+\left [\frac{27}{8}-\frac{19}{8}\eta+\frac{1}{24}\eta^{2}
+\left (\frac{23}{12}-\frac{31}{6}\eta-\frac{1}{4}\eta^{2}\right )e^{2}
+\left (\frac{1}{24}-\frac{35}{24}\eta
-\frac{1}{8}\eta^2\right )e^{4}\right ]\zeta^{2}
\nonumber\\
&&
+\left [\frac{135}{16}-\left (\frac{6889}{144}-\frac{41}{24}\pi^{2}
\right )\eta
+\frac{31}{24}\eta^{2}
+\frac{7}{1296}\eta^{3}
\right.
\nonumber\\
&& 
+\left (\frac{299}{16}-\left (\frac{10003}{144}-\frac{41}{24}\pi^{2}
\right)\eta
+\frac{3013}{216}\eta^{2}-\frac{5}{144}\eta^{3}\right )e^{2}
\nonumber\\
&& 
\left.\left. 
+ \left (\frac{77}{144}-\frac{6497}{432}\eta+\frac{853}{72}\eta^{2}
+\frac{5}{16}\eta^{3}\right)e^{4}
\right.
\right.
\nonumber\\
&&
\left.
\left.
-\left (\frac{7}{1296}+\frac{1}{16}\eta+\frac{1}{8}\eta^{2}
-\frac{1}{16}\eta^{3}\right )e^{6}
\right ]\zeta^{3}\right \} \,.
\label{Jharm}
\end{eqnarray}
\label{EJharm}
\end{subequations}
Notice that $E_{\rm Harm}$ is proportional to $(1-e^2)$ through 3PN order,
indicating that $E_{\rm Harm}=0$ for the limiting unbound orbit $e=1$; this
is another appropriate feature of our ``covariant'' eccentricity.
The energy and angular momentum 
are well-defined, physically observable quantities, 
so one can alternatively express our orbit elements $\zeta$ and $e$ 
as functions of $\tilde{E}$ and $\tilde{J}$. 
Here we give the results to 1PN order, but the calculation can be done 
to 3PN order:
\begin{eqnarray}
\zeta&=&\frac{1}{\tilde{J}^{2}}
\left [1+\frac{2}{3\tilde{J}^{2}}\left(
4+2\eta-(1-3\eta)\tilde{E}\tilde{J}^{2}\right)\right],
\nonumber \\
e&=&\sqrt{1+2\tilde{E}\tilde{J}^{2}}
\left [1-\frac{1}{2}\frac{\tilde{E}}{1+2\tilde{E}\tilde{J}^{2}}\left (4+2\eta-(1-3\eta)\tilde{E}\tilde{J}^{2}\right)\right].
\label{euEJ}
\end{eqnarray}

\subsection{ADM vs. Harmonic gauge}\label{gauge}

The foregoing results are valid in harmonic gauge.  That gauge is
characterized by the condition $\partial_{\nu}h^{\mu\nu}=0$,
where
$h^{\mu\nu}=\sqrt{-g}g^{\mu\nu}-\eta^{\mu\nu}$,
$g_{\mu\nu}$ and $g$ are the physical metric and its determinant, and
$\eta_{\mu\nu}$ is a background Minkowski metric.  In this gauge,
Einstein's equations take the form
\begin{equation}
\Box h_{\mu\nu}=16\pi\tau_{\mu\nu},
\end{equation}
where $\square=\eta^{\mu\nu}\partial_{\mu}\partial_{\nu}$ 
is the flat d'Alembertian operator, and the source term 
$\tau_{\mu\nu}$ 
depends both on the matter stress-energy tensor $T^{\mu\nu}$ 
and on non-linear contributions
of the gravitational field.  The local equation of motion $\nabla_\nu
T^{\mu\nu}=0$ is equivalent to $\partial_\nu \tau^{\mu\nu}=0$, which
follows from the harmonic gauge condition.  
There actually is an infinity of distinct harmonic gauges, and 
the equations of motion will generally depend on the choice of a 
particular gauge.  We already saw an example of this in the choice of
eliminating logarithmic terms from the 3PN contributions [Eq.
(\ref{gaugelog}) above].  

A different approach to the two-body problem,  implemented through 3PN 
order by Damour, Jaranowski and Sch\"afer 
\cite{jaraschafer98,jaraschafer99,djs00}, 
is to compute the Hamiltonian of the system rather than the equations 
of motion. Unlike other methods, this does not use a harmonic 
coordinate system, but a so-called ADM (Arnowitt-Deser-Misner), 
or ``Hamiltonian'' gauge, or coordinate system.
It 
has been proven to be equivalent to the harmonic formulation \cite{abf01}. 

The Hamiltonian has been computed up to 3PN order; because it is a
Hamiltionian approach, it explicitly suppresses the 2.5PN and 3.5PN
contributions of radiation reaction. 
We quote it here only to 1PN order 
(see \cite{djs00} for a complete expression):
\begin{eqnarray}
H_{\textrm{ADM}} & = & \frac{p_{1}^{2}}{2m_{1}}-\frac{m_{1}m_{2}}{2r}
+\left [-\frac{p_{1}^4}{8m_{1}^{3}}+\frac{m_{1}^{2}m_{2}}{2r^2}
\right. 
\nonumber\\ 
& &\left. 
+\frac{m_{1}m_{2}}{r}\left (
\frac{(\mathbf{n}\cdot\mathbf{p_{1}})(\mathbf{n}\cdot\mathbf{p_{2}})}
{4m_{1}m_{2}}
-\frac{3}{2}\frac{p_{1}^{2}}{m_{1}^{2}}
+\frac{7}{4}\frac{\mathbf{p_{1}}\cdot\mathbf{p_{2}}}{m_{1}m_{2}}
\right )\right ]
\nonumber\\ 
& &
+ (1\rightleftharpoons 2) \,.\label{hadm}
\end{eqnarray}
We convert this two-body problem into an effective one-body problem 
by using the simple relation $\mathbf{p}=\mathbf{p_{1}}=-\mathbf{p_{2}}$, 
valid in the center-of-mass frame. 
Thus we get a new expression for $H_{\textrm{ADM}}(\mathbf{x},\mathbf{p})$.

From Hamilton's equations:
\begin{equation}
\frac{d \mathbf{x}}{d t}=\nabla_{\mathbf{p}}H_{\textrm{ADM}}\,,
\qquad\frac{d \mathbf{p}}{d t}=-\nabla_{\mathbf{x}}H_{\textrm{ADM}},
\end{equation}
we iteratively extract the equations of motion and write them in 
the same form as equation (\ref{eom}), but
with different $A$ and $B$. Substituting 
the expression of $\mathbf{p}$ as a function of $\mathbf{v}$ 
and $\mathbf{x}$ into the Hamiltonian, we obtain the total conserved 
energy $E_{\textrm{ADM}}$. Similarly, we get $\mathbf{J}_{\textrm{ADM}}$ 
by calculating $\mathbf{x}\times\mathbf{p}$. 
For both the equations of motion and the 
expressions for energy and angular momentum, the
harmonic and ADM-Hamiltonian terms coincide at 1PN order, 
but they differ at 2PN and 3PN orders. 

We apply the method described in section \ref{osculate} 
to find solutions to the ADM equations of motion and
expressions for $E_{\textrm{ADM}}$ and $\mathbf{J}_{\textrm{ADM}}$
in terms of the osculating orbit elements. 
In this case, $\tilde{e}$ and $\tilde{u}$ are strictly constant
because radiation reaction is not present in the Hamiltonian
approach.  We then find expressions for our new orbit elements
${e}$ and ${\zeta}$ in terms of  $\tilde{e}$ and $\tilde{u}$ 
and write $E_{\textrm{ADM}}$ and $\mathbf{J}_{\textrm{ADM}}$
in terms of these elements.  The results are:
\begin{subequations}
\begin{eqnarray}
\label{EADM}
\tilde{E}_{\textrm{ADM}}&=&-\frac{1}{2}(1-e^{2})\zeta
\left \{ 1 -\left [ \frac{3}{4}+\frac{1}{12}\eta-\left (\frac{1}{12}-\frac{1}{4}\eta\right )e^{2}
\right ]\zeta
\right .
\nonumber\\
 &&
 \left .
 -\left [\frac{27}{8}-\frac{19}{8}\eta+\frac{1}{24}\eta^{2}
-\left (\frac{17}{12}+\frac{7}{4}\eta+\frac{1}{4}\eta^{2}\right )e^{2}
+\left (\frac{1}{24}+\frac{11}{24}\eta-\frac{1}{8}\eta^{2}\right )e^{4}
\right ]\zeta^{2}
\right.
\nonumber\\
&&
\left.
-\left [ \frac{675}{64}-\left (\frac{34445}{576}-\frac{205}{96}\pi^{2} 
\right)\eta
+\frac{155}{96}\eta^{2}+\frac{35}{5184}\eta^{3}
\right.
\right.
\nonumber\\
&&
\left.
\left.
+\left (\frac{7}{64}+\left (\frac{167}{64}-\frac{41}{96}\pi^{2}\right )\eta
+\frac{7595}{864}\eta^{2}-\frac{25}{576}\eta^{3}\right )e^{2}
\right.
\right.
\nonumber\\
&&
\left.
\left.
-\left (\frac{815}{576}-\frac{6995}{1728} \eta
-\frac{299}{288}\eta^{2}-\frac{25}{64}\eta^{3}\right )e^{4}
\right.
\right.
\nonumber\\
&&
\left.
\left.
-\left(\frac{35}{5184}-\frac{31}{192}\eta+\frac{13}{32}\eta^{2}
-\frac{5}{64}\eta^{3}\right )e^{6}
\right ]\zeta^{3} \right \} \,,
\\
\label{JADM}
\tilde{J}_{\textrm{ADM}}&=&\frac{1}{\sqrt{\zeta}} \left \{ 1
+\left[ \frac{3}{2}+\frac{1}{6}\eta
-\left (\frac{1}{6}-\frac{1}{2}\eta\right )e^{2}\right ]\zeta
\right .
\nonumber\\
&&
\left .
+\left [\frac{27}{8}-\frac{19}{8}\eta+\frac{1}{24}\eta^{2}
+\left (\frac{23}{12}-\frac{35}{12}\eta-\frac{1}{4}\eta^{2}\right
)e^{2}
+\left (\frac{1}{24} -\frac{17}{24}\eta
-\frac{1}{8}\eta^2\right )e^{4}\right ]\zeta^2
\right .
\nonumber\\
&&
\left .
+\left [\frac{135}{16}-\left (\frac{6889}{144}
-\frac{41}{24}\pi^{2} \right )\eta
+\frac{31}{24}\eta^{2}
+\frac{7}{1296}\eta^{3}
\right.\right.
\nonumber\\
&&\left.\left.
+\left (\frac{299}{16}
-\left (\frac{1025}{16}
-\frac{41}{24}\pi^{2} \right)\eta
+\frac{2077}{216}\eta^{2}-\frac{5}{144}\eta^{3}\right )e^{2}
\right.\right.
\nonumber\\
&&\left.\left.
+ \left (\frac{77}{144}-\frac{1337}{432}\eta
+\frac{271}{72}\eta^{2}+\frac{5}{16}\eta^{3}\right)e^{4}
\right.
\right.
\nonumber\\
&&
\left.
\left.
-\left (\frac{7}{1296}-\frac{7}{48}\eta+\frac{3}{8}\eta^{2}
-\frac{1}{16}\eta^{3}\right )e^{6}
\right ]\zeta^3 \right \} \,.
\nonumber \\
\end{eqnarray}
\label{EJadm}
\end{subequations}

We observe two features of the 
harmonic and the ADM versions of these expressions: 
(i) the ``circular'' parts ($e=0$) of the formulae coincide. In that 
case the angular velocity $\Omega=\Omega_{a}=\Omega_{p}$ 
is the same as that observed from infinity for both harmonic 
and ADM coordinates; (ii) the expressions also coincide for 
$\eta\rightarrow 0$, \emph{i.e.} in the test-mass limit. 
As mentioned before, the differences between the formulae only occur at 
2PN and 3PN orders.  It is actually possible to relate the coordinate 
positions and velocities in the two gauges.  In particular, the 
relation between $\dot{\phi}_{\textrm{ADM}}$ 
and $\dot{\phi}_{\textrm{Harm}}$, ${r}_{\textrm{Harm}}$, 
\emph{etc.} allows us to find a relation 
between $({e}_{\textrm{ADM}},{\zeta}_{\textrm{ADM}})$ 
and $({e}_{\textrm{Harm}},{\zeta}_{\textrm{Harm}})$, and thus account 
for the differences in the coefficients of $E$ and $\mathbf{J}$. 
We found that a transformation of the type
\begin{eqnarray}
\dot{\phi}_{\textrm{ADM}}&=&
\dot{\phi}\left \{1+\eta\frac{m}{r}\left [
\frac{9}{4}\left (v^{2}-\frac{m}{r}\right )
-\left (\frac{16}{3}+\frac{\eta}{2}\right )
{\left (\frac{m}{r}\right )}^{2}
\right.\right.
\nonumber\\ 
&&
\left.\left.
+\left (\frac{17}{8}-\frac{21}{4}\eta\right )v^{4}
+\left (\frac{239}{24}+\frac{7}{2}\eta\right )\frac{m}{r}v^{2}
+\dot{r}^{2}f\left(\dot{r}^{2},\frac{m}{r},v^{2}\right)\right ]\right \},
\label{2phis}
\end{eqnarray}
where we have dropped the subscript ``Harm'' in the right-hand-side of Eq.
(\ref{2phis}),
and where $f$ is a function, was 
compatible with the differences observed in the expressions of 
both the energy and the angular momentum.  Since $\dot{r}=0$ at 
the apastron and periastron, $f$ does not 
need to be determined explicitly for our purposes.  
In the circular orbit limit, where, from
Eq. (\ref{eom}), $v^2=m/r[(1-(3-\eta)m/r]$ to PN order, it is easy to see
that $\dot{\phi}_{\textrm{ADM}}=\dot{\phi}_{\textrm{Harm}}$.  Eq.
(\ref{2phis}) demonstrates that our definitions of $e$ and $\zeta$ are not
truly covariant.  Nevertheless, the coordinate transformations
that connect different formulations of the post-Newtonian equations of motion
cause changes beginning only at
2PN order.   This is reflected in Eq. (\ref{2phis}) where the difference
between the two angular velocities is of 2PN order.  Furthermore, for the
small eccentricity orbits that we wish to consider, the corrections 
are proportional to $e$, and are thus further suppressed.  Thus we argue
that our definitions of $e$ and $\zeta$ are ``almost'' covariant.

\section{Effects of finite size}
\label{finitesize}

\subsection{Estimates for compact binaries}

In reality, the bodies in our binary system cannot be treated as purely
point masses.  They may be rotating, and thus subject
to a number of effects, including rotational kinetic energy, rotational
flattening, and spin-orbit and spin-spin interactions.  Furthermore,
there will be tidal deformations.
These effects will not only make direct contributions to the energy
and angular momentum of the system, they may also modify the equations
of motion, and thereby modify the expressions for our alternative 
eccentricity and
semi-latus rectum.
However because they depend on the size of the bodies, which, for
neutron stars and black holes, are of order $m$, we expect these
effects to be ``effectively'' of high PN order, even if they are
Newtonian in origin, such as tidal effects.  To see this, we estimate
each finite-size effect in turn and compare it with the 
Newtonian orbital energy $E_N \sim m^2/r$.  We assume that the rotational
angular velocity $\omega$ of each body ranges from zero to the orbital angular
velocity, given by $\Omega \sim (m/r^3)^{1/2}$, 
and we let the radius of each body
be of the form $R_a \sim q m_a$, where $q \sim 1$ for black
holes (in harmonic coordinates), and $q \sim 5$ for neutron
stars.  

\begin{itemize}

\item
Rotational kinetic energy:  $E_{\rm Rot} \sim I \omega^2/2 \le mR^2(m/r^3)
\sim E_N q^2 (m/r)^2$.  This is effectively 2PN order.  There will be
PN corrections to the kinetic energy, given by $E_{\rm Rot-PN} \sim E_{\rm
Rot} (R\omega)^2 \sim mR^4\omega^4 \sim E_N q^4 (m/r)^5$.  These are
effectively 5PN order, but, because of the $q^4$ dependence,
could be important for neutron stars.

\item
Rotational flattening: $E_{\rm Flat} \sim \delta I \omega^2/2$, where
$\delta$ is a measure of the deformation of the body, given by the
ratio of rotational to gravitational energy, $\delta \sim
(I\omega^2)/(m^2/R)$, so that $E_{\rm Flat} \sim \omega^4 R^5 \le
E_N q^5 (m/r)^5$.  There is an equivalent contribution of
rotational flattening to the gravitational internal energy.
These are effectively 5PN order, but because
of the $q^5$ dependence, could be important for neutron stars.

\item
Tidal deformations: $E_{\rm Tidal} \sim (\delta^\prime m)^2 /R$, where 
$\delta^\prime$ is the ratio of gravitational energy due to the tidal
force of the companion to the internal gravitational energy of the
body, $\delta^\prime \sim (mR^2/r^3)/(m/R) \sim (R/r)^3$.  Thus
$E_{\rm Tidal} \sim m^2 (R/r)^6/R \sim E_N q^5 (m/r)^5$.
There is also a contribution from the rotational kinetic energy of the
tidal bulge, given by $E_{\rm KE-bulge} \sim \delta^\prime I\omega^2
\sim mR^5\omega^2/r^3 \le E_N q^5 (m/r)^5$.  These are
effectively 5PN order, but could be significant for neutron stars.

\item
Spin-orbit coupling: $E_{\rm S.O.} \sim LS/r^3 \sim
(mr^2\Omega)(mR^2\omega)/r^3 \le E_N q^2 (m/r)^3$.  This is
effectively 3PN order \cite{spincomment}, and generally must be included.

\item
Spin-spin coupling: $E_{\rm S.S.} \sim S_1 S_2/r^3 \sim
(mR^2\omega)^2/r^3  \le E_N q^4 (m/r)^5$.  This is effectively
5PN order, but could be significant for neutron stars \cite{spincomment}.

\end{itemize}

A parallel heirarchy of finite-size 
effects applies to the total angular momentum of
the system.

The largest effect in principle is that due to the rotational kinetic
energy of the bodies and thus requires some care.  For black holes, we can
apply the general formulas for mass and angular momentum of
isolated  Kerr black holes, in terms of the irreducible mass and
angular velocity.  For neutron stars, no such general formula exists,
so it may be necessary to rely upon numerical results for energy and
angular momentum of isolated rotating neutron star models in order to take
accurate account of this effect.   On the other hand, it does not directly
affect the equations of motion.
 
Because the remaining effects are effectively of 3PN order and higher, 
our strategy
will be to evaluate them analytically to the lowest non-trivial order.
For tidal and rotational flattening terms, this will mean using
Newtonian theory.  For spin-orbit and spin-spin terms, we will use the
well-known 1PN formulae.  
We will ignore any coupling among these effects, or between these effects
and the point-mass PN effects described in the previous section.  Accordingly,
we will
calculate the separate
contribution of each effect to the energy and angular momentum
and simply add them all up.

\subsection{Newtonian Tidal and Rotational Effects}

In Appendix \ref{appa} we derived the general form of the equations of
motion and the conserved energy and angular momentum for a binary system
of
tidally and rotationally deformed bodies, and in Appendix \ref{appb} we
specialized to linear perturbations and multipole indices $l=2$ and $l=3$. 
We  now specialize further to systems more relevant to the initial
configurations in numerical relativity which we wish to study, namely 
binary systems in which the spin axes of both stars are
perpendicular to the orbital plane.  The equation of motion (\ref{eomnew})
then takes the simplified form
\begin{equation}
\mathbf{a} = - \frac{m}{r^2}\mathbf{n} 
\left [ 1 + A \left ( \frac{m}{r} \right )^2
+ B \left ( \frac{m}{r} \right )^5 +
C \left ( \frac{m}{r} \right )^7 \right ] \,,
\end{equation}
where the three perturbing terms correspond repectively to the effects of
rotational distortions, quadrupole tidal distortions ($l=2$) and octupole
tidal distortions ($l=3$), with the coefficients given by
\begin{eqnarray}
A &=& m^{-2} ( R_1^5 k_2^{(1)} \tilde{\omega}_1^2/m_1 + R_2^5 k_2^{(2)}
\tilde{\omega}_2^2/m_2) \,,
\nonumber \\
B &=&  6m^{-5} ( R_1^5 k_2^{(1)} m_2/m_1 + R_2^5 k_2^{(2)} m_1/m_2) \,,
\nonumber \\
C  &=&  8m^{-7} ( R_1^7 k_3^{(1)} m_2/m_1 +  R_2^7 k_3^{(2)} m_1/m_2) \,.
\end{eqnarray}
For each body,
$R_a$ denotes its radius, $k_2^{(a)}$ and $k_3^{(a)}$
denote the ``apsidal constants'' for angular harmonics 
$l=2$ and $l=3$,
respectively, and $\tilde{\omega}_a$ denotes the body's angular velocity 
at a chosen point in the orbit (see Appendix \ref{appb} for details).
Apsidal constants are dimensionless coefficients that depend on the degree
of central condensation of the star, and that determine the size of 
distortion of a given angular degree $l$ produced by a given external
perturbation.  
Note that $A < R^5/m^2r^3 \sim q^5 (m/r)^3$, so that, despite appearances,
this term, like the purely tidal term from $B$, is effectively 5PN order.
The energy and angular momentum that are conserved by virtue of the full
fluid equations of motion are given by
\begin{eqnarray}
E &=& E_{\rm Self} + E_{\rm Distort} + E_{\rm TR,Orbit} 
\nonumber \\
 &=& \left [ 
 \frac{1}{2} I_1 \tilde{\omega}_1^2
 - {\cal W}_1  + (1\rightleftharpoons 2) \right ]
 + \left [ \frac{1}{3} R_1^5 k_2^{(1)} \tilde{\omega}_1^2 \left ( 
 \tilde{\omega}_1^2 + 2 \frac{m_2}{\tilde{r}^3} \right )
 + (1\rightleftharpoons 2) \right ]
   \nonumber \\
   &&
 +\frac{1}{2} \mu v^2 - \frac{\mu m}{r} 
     \left [1 + 
     \frac{1}{3} A \left ( \frac{m}{r} \right )^2
     + \frac{1}{6} B \left ( \frac{m}{r} \right )^5 
     + \frac{1}{8} C \left ( \frac{m}{r} \right )^7 \right ] \,,
\nonumber \\
J &=&  
S + J_{\rm Distort} +J_{\rm TR,Orbit} 
\nonumber \\
&=& 
 \left [ I_1 \tilde{\omega}_1 +  I_2 \tilde{\omega}_2 \right ]
 + \left [  \frac{2}{3} R_1^5 k_2^{(1)} \tilde{\omega}_1 \left (
 \frac{2}{3} \tilde{\omega}_1^2 
  + \frac{m_2}{\tilde{r}^3} \right )
 +  ( 1\rightleftharpoons 2 )
 \right ]
    \nonumber \\
       &&
      + \mu |{\bf x} \times {\bf v}|
      \,,
\label{EJtidal}
\end{eqnarray}
where, for each body, $I_a$ denotes the moment of inertia,
${\cal W}_a$ denotes the self-gravitational energy of the undistorted
configuration, and $\tilde{r}$ denotes
the orbital separation at the point at which the star's angular velocity is
$\tilde{\omega}$.
The chosen point in our case will be the pericenter or apocenter.  In Eq.
(\ref{EJtidal}),
the split among the intrinsic energy and spins of the bodies 
$E_{\rm Self}$ and $S$, the constant distortion terms $E_{\rm Distort}$ and
$J_{\rm Distort}$, and the orbital terms 
is
clear.  The angular momentum components are all referred to the axis
perpendicular to the orbital plane.

We now repeat the method of subsections \ref{osculate} -- \ref{conserved} to
obtain the general solution to the equations of motion
to first order in the tidal and rotational perturbations.  We then
obtain our new orbit elements $e$ and $\zeta$ in terms 
of the bare elements $\tilde{e}$ and $\tilde{u}$; for example, $e$ is given by
\begin{eqnarray}
e &=& \tilde{e} \left \{ 
1 - \frac{1}{2} \left [ 1 + \frac{2}{3} \tilde{e}^2 \right ] A \tilde{u}^2
+ \frac{1}{4} \left [1 - \frac{85}{12} \tilde{e}^2 -\frac{85}{24}
\tilde{e}^4 \right ] B \tilde{u}^5
\right .
\nonumber \\
&&
\left .
 + \frac{1}{4} \left [3 - \frac{49}{8}\tilde{e}^2 -\frac{147}{8}\tilde{e}^4
-\frac{931}{256}\tilde{e}^6 \right ] C \tilde{u}^7
\right \} \,.
\label{etidal}
\end{eqnarray}
Since we are assuming that these effects are {\em effectively} of 5PN 
order, we can simply add the correction terms in Eq.
(\ref{etidal}) to those in Eq. (\ref{e3PN}).
Tidal and rotational interactions are conservative (as long as we
ignore dissipative processes such as viscosity), and therefore do
not cause secular evolution of $\tilde{e}$ or $\tilde{u}$; however
they do produce a pericenter advance, given in terms of our new orbit
elements by
\begin{equation}
\frac{d\omega}{d\phi} = 
A \zeta^2
+ \frac{5}{2} \left [ 1 + \frac{3}{2} e^2 +
\frac{1}{8} e^4 \right ] B\zeta^5
+ \frac{7}{2} \left [ 1+ \frac{15}{4} e^2 + \frac{15}{8} e^4
  + \frac{5}{64} e^6 \right ] C \zeta^7 \,.
\end{equation}

Substituting the solutions for the motion into the orbital parts of 
Eqs. (\ref{EJtidal}), and converting to our new elements, we obtain
for the tidal-rotational (TR) contributions to the orbital parts of 
$E$ and $J$, 
\begin{subequations}
\begin{eqnarray}
{E}_{\rm TR,Orbit} &=&
\mu (1-e^2) \left [
\frac{1}{9} (3- e^2) A \zeta^3
+ \frac{1}{18}( 9+ 10 e^2 - 3e^4) B \zeta^6
\right .
\nonumber \\
&& 
\left .
+ \frac{1}{24} (13 + 49e^2 + 7e^4 -5e^6 ) C \zeta^8
\right ]
\,,
\label{ETRorbit}
\\
{J}_{\rm TR,Orbit} &=&
\mu m \left [ \frac{2}{9} (3 + e^2 ) A \zeta^{3/2}
+ \frac{2}{9} (3+ 10e^2 + 3e^4 ) B \zeta^{9/2}
\right .
\nonumber \\
&& 
\left .
+ \frac{2}{3} (1 +7e^2 + 7e^4 + e^6 ) C \zeta^{13/2} \right ]
\,.
\label{JTRorbit}
\end{eqnarray}
\label{EJTRorbit}
\end{subequations}
where we have dropped the Newtonian orbital part, because it is already
included in the 3PN point-mass expressions of Eqs. (\ref{EJharm}) or
(\ref{EJadm}).  
The form of the self-terms depends on where in the orbit we evaluate the
stars' angular velocities; for pericenter or apocenter, we can use the
Newtonian relation that $m/\tilde{r} = \zeta (1 \pm e)$, respectively, to
write
\begin{subequations}
\begin{eqnarray}
{E}_{\rm Self} &=&
\frac{1}{2} I_1 \tilde{\omega}_1^2
 - {\cal W}_1  + (1\rightleftharpoons 2) \,,
 \label{Efinalself}
\\
S &=&
 I_1 \tilde{\omega}_1 + I_2 \tilde{\omega}_2 \,,
 \label{Sfinal}
 \\
 {E}_{\rm Distort}  &=&
  \frac{1}{3} m^{-2}R_1^5 k_2^{(1)} \tilde{\omega}_1^2 \left [
   (m \tilde{\omega}_1)^2 + 2 \frac{m_2}{m} \zeta^3 (1 \pm e)^3 \right ]
    + (1\rightleftharpoons 2)
\,,
\label{Efinaldistort}
 \\
 J_{\rm Distort}  &=&
 \frac{2}{9} m^{-2}R_1^5 k_2^{(1)}
 \tilde{\omega}_1\left [
    2 (m \tilde{\omega}_1)^2 +  
    3 \frac{m_2}{m} \zeta^3 (1 \pm e)^3 \right ]
	+ (1\rightleftharpoons 2)
	\,.
	\label{Jfinaldistort}
\end{eqnarray}
\label{EJself}
\end{subequations}

\subsection{Spin-orbit and Spin-Spin effects}

Spin-orbit and spin-spin interactions produce corrections in the equations
of motion that are formally of 1PN order.  For systems with the spins
perpendicular to the orbital plane they are given by
\begin{eqnarray}
{\bf a} = -\frac{m}{r^2}{\bf n} &+& \frac{1}{r^3} 
\left \{ 6{\bf n} ({\bf n} \times
{\bf v}) \cdot \left ( 2{\bf S} + \frac{\delta m}{m} \fourvec{\Delta} \right )
- {\bf v}\times \left ( 7 {\bf S} + 3\frac{\delta m}{m} \fourvec{\Delta}
\right )
\right .
\nonumber \\
&& 
\left .
+3 \dot{r} {\bf n} \times \left ( 3 {\bf S} + \frac{\delta m}{m}
\fourvec{\Delta} \right ) 
\right \}
- \frac{3}{\mu r^4} {\bf n} {\bf S}_1 \cdot {\bf S}_2 \,,
\end{eqnarray}
where ${\bf S}={\bf S}_1+{\bf S}_2$ and  $\fourvec{\Delta}=m({\bf S}_2/m_2 -
{\bf S}_1/m_1 )$.  The individual spins are constants of the motion
when they are both aligned perpendicular to the orbital plane.
The conserved energy and total angular momentum are given
by
\begin{subequations}
\begin{eqnarray}
E &=& \frac{1}{2} \mu v^2 - \frac{\mu m}{r} +
\frac{1}{r^3} {\bf L}_N \cdot \left ( {\bf S} + \frac{\delta m}{m}
\fourvec{\Delta} \right )
- \frac{1}{r^3} {\bf S}_1 \cdot {\bf S}_2 \,,
\label{Espin}
\\
J &=& {L}_N+ {S}- \eta \left [ \frac{m}{r} \left ( 3 {S}
+ \frac{\delta m}{m}
{\Delta} \right ) - \frac{1}{2} v^2 \left ( {S} 
+ \frac{\delta m}{m} {\Delta} \right ) \right ]
\,,
\label{Jspin}
\end{eqnarray}
\label{EJspin}
\end{subequations}
where ${\bf L}_N = \mu {\bf x} \times {\bf v}$, 
and Eq. (\ref{Jspin}) denotes
the component perpendicular to the orbital plane
(for the complete equations of motion,
see, for example, \cite{kww,kidder}).
We define the
dimensionless quantities
\begin{equation}
D \equiv \eta \frac{S}{L_N} \,, \quad F \equiv \eta \frac{\delta m}{m}
\frac{\Delta}{L_N} \,, \quad G\equiv \eta \frac{S_1 S_2}{(L_N)^2}
\,,
\end{equation}
where $D \sim F \sim (R/r)^2 \sim q^2 (m/r)^2$, and
$G \sim (R/r)^4 \sim q^4 (m/r)^4$, making the spin-orbit and spin-spin
terms effectively 3PN and 5PN order respectively \cite{spincomment}.
With these definitions, the equation of motion takes the form of Eq.
(\ref{eom}), with
\begin{eqnarray}
A &=& (5D+3F-3G)v^2 - 3(D+F-G)\dot{r}^2 \,,
\\
B &=& -2D \dot{r}  \,.
\end{eqnarray}
Again we solve the equations of motion using the method of subsections
\ref{osculate} -- \ref{conserved} and define our new orbit elements.  In
this case, for example, the eccentricity is given by 
\begin{equation}
e = \tilde{e} \left \{ 1 + \frac{1}{2} \left [ (1+4\tilde{e}^2)D +
(3+2\tilde{e}^2)(F-G) \right ] \tilde{u} \right \}
\,.
\end{equation}
In terms of our new elements, the pericenter advance is given by
\begin{equation}
\frac{d\omega}{d\phi} = - (7D+3F-3G)\zeta \,.
\end{equation}
while $e$ and $\zeta$ undergo no secular changes. 
When expressed in terms of our new orbit
elements, the spin-orbit and spin-spin contributions to the total energy
and angular momentum
have the form
\begin{subequations}
\begin{eqnarray}
{E}_{\rm Spin} &=& - \frac{1}{3} \mu (1-e^2) \left [ (7-2e^2) D +
(3-e^2)(F-G)
\right ] \zeta^2 \,,
\label{Espinfinal}
\\
{J}_{\rm Spin} &=& - \frac{1}{6} \mu m \left [ 5(7+e^2)D + (15+e^2)F -
4(3+e^2) G \right ] \sqrt{\zeta}
\,.
\label{Jspinfinal}
\end{eqnarray}
\label{EJspinfinal}
\end{subequations}
Inserting the Newtonian expression for $L_N$, we have that
\begin{equation}
D = (S/m^2)\sqrt{\zeta} \,, \quad
F = (\delta m /m)(\Delta/m^2)\sqrt{\zeta} \,, \quad
G = (S_1/m_1)(S_2/m_2)m^{-2} \zeta \,.
\end{equation}

\subsection{Other finite-size corrections}

In deriving the ``point-mass'' equations of motion, the underlying
assumption was that the masses that enter the equations are the total
mass of each body, comprised of
baryonic mass, gravitational binding energy and rotational
kinetic energy, if any.  
Thus, each $m_a$ should be written $m_a = m_a^{\rm B} - {\cal
W}_a + E_a^{\rm Rot}$.  In many numerical
approaches, sequences of models are constructed in which the total (or
ADM) mass of each corresponding non-rotating star  is held fixed along the
sequence.  Thus, for making comparisons with such sequences, we should
replace each $m_a$ in Eqs. (\ref{EJharm}) or (\ref{EJadm})
with $m_a^0 + I_a \omega_a^2 /2$ (or, in the case of black holes, with a
suitable formula in terms of the irreducible mass and $\omega_a$).  But
because $I_a \omega_a^2 \sim q^2 (m/r)^3$, the main contribution, at
effectively 3PN order, comes from making this replacement in the Newtonian
expressions.  Expressing $E_N$ and $J_N$
in terms of $\Omega$ as $E_N = -\frac{1}{2} \eta m
(1-e^2) (m\Omega_a)^{2/3}/(1-e)^{4/3}$, and $J_N = \eta m^2
(1-e)^{2/3}/(m\Omega_a)^{1/3}$, and making the above replacement, we
find the corrections to the Newtonian energy and angular momentum
\begin{subequations}
\begin{eqnarray}
{E}_{\rm N,Corr} &=& 
-\frac{1}{4} \mu (1-e^2) \zeta \left [ \frac{I_1}{m_1} \omega_1^2 
\left (1-\frac{m_1}{3m} \right ) + (1 \rightleftharpoons 2 )
\right ] \,,
\\
{J}_{\rm N,Corr} &=&
\frac{\mu m}{2\sqrt{\zeta}} \left [ \frac{I_1}{m_1} \omega_1^2
\left (1-\frac{m_1}{3m} \right ) + (1 \rightleftharpoons 2 )
\right ] \,,
\end{eqnarray}
\label{EJcorr}
\end{subequations}
where all masses now are those of the equivalent non-rotating body.  For
neutron stars, this would be that of the same baryonic mass; for black
holes, it would be that of the same irreducible mass.

\section{A post-Newtonian diagnostic for quasi-equilibrium configurations}
\label{pndiagnostic}

\subsection{Estimates of effects}

We now have all the ingredients to formulate a post-Newtonian diagnostic for
quasi-equilibrium configurations of compact bodies.  
The ingredients are the various contributions to the total energy and
angular momentum of the system in terms of the ``covariant'' 
orbit elements 
$e$ and $\zeta$, together with the relationships connecting the value of
$\zeta$ with the orbital angular velocity at a turning point of the orbit,
namely $\zeta=(m\Omega_p)^{2/3}/(1+e)^{4/3}$ or
$\zeta=(m\Omega_a)^{2/3}/(1-e)^{4/3}$, corresponding to pericenter and
apocenter, respectively.  The ingredients are:

\begin{itemize}
\item
{\em Point-mass orbital contributions through 3PN order}.  Eqs.
(\ref{EJharm}) or (\ref{EJadm}).  It is straightforward to show that,
because the harmonic and ADM
versions differ by 2PN terms proportional to $\eta e^2$ and
higher, the differences between the two versions are negligible for all
cases of interest.  Henceforth we will adopt the harmonic version of Eqs.
(\ref{EJharm}).

\item
{\em Self Terms}.  Eqs. (\ref{Efinalself}) and (\ref{Sfinal}).  
We add a suitably defined total 
``rest'' mass for the bodies to the definition of
$E_{\rm Self}$. 
Because the rotational kinetic energy and the spin angular momentum
are effectively of 2PN
order, they will have to be treated with some care.

\item
{\em Constant distortion terms}.  Eqs. (\ref{Efinaldistort}) and
(\ref{Jfinaldistort}).

\item
{\em Tidal-rotational orbit terms}. Eqs. (\ref{EJTRorbit}) .

\item
{\em Spin-orbit and Spin-spin terms}.   Eqs. (\ref{EJspinfinal}).

\item
{\em Newtonian correction terms}.   Eqs. (\ref{EJcorr}).

\end{itemize}

In order to assess the applicability of this diagnostic, we first
study the sizes of various effects for systems of interest.  In general we
will consider systems of solar-mass scale neutron stars or black holes, in
circular or small-eccentricity orbits, in the vicinity of the onset of an
unstable plunge and  merger.  This corresponds to $\zeta
\sim m/r < 1/5$ for black holes, or to $\zeta < m/(2R) \sim 1/q$ for
neutron stars.   For $q$ between 4 and 6, the two ranges are
comparable.
Both correspond to $m\Omega < 0.1$.  We will generally choose a range
$0.01 < m\Omega < 0.1$.  

\begin{figure}
\leavevmode
\psfig{figure=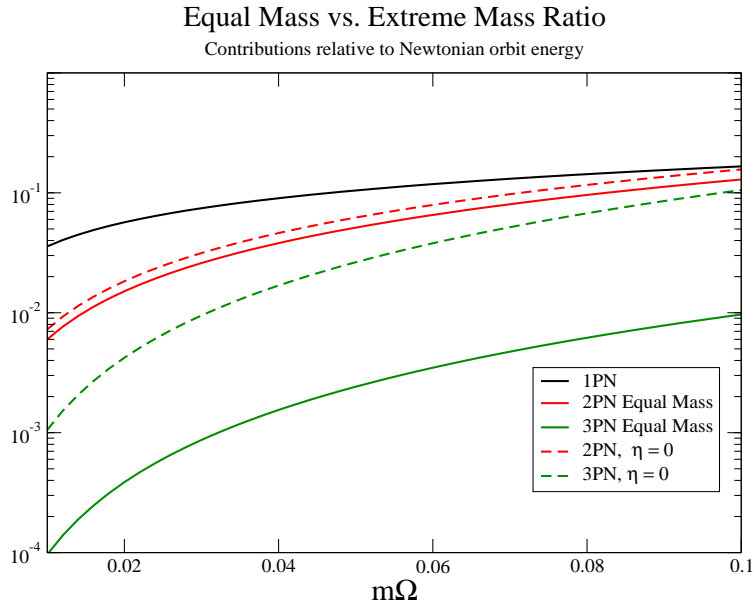,angle=270,width=12cm}
\caption{\label{Fig1} Contributions of 1PN, 2PN and 3PN terms to the energy,
expressed as a fraction of the Newtonian energy, vs. $m\Omega$.  
Circular orbits are assumed. 
Shown are the equal mass
case ($\eta=1/4$) and the point-mass limit ($\eta=0$).}
\end{figure}

First we look at the relative contributions of point-mass PN corrections.
Figure 1 shows the contribution, relative to the Newtonian orbital 
energy, of the
1PN, 2PN and 3PN terms in the energy, for $\eta =0$ and $\eta =1/4$, as a
function of $m\Omega$.  Results for the angular momentum are similar.
While the 1PN terms are essentially insensitive to
$\eta$, and the 2PN terms are only 15 \% smaller for equal masses than
for the test-mass limit, the 3PN terms are suppressed for equal masses
by more than a factor
of 10 compared to the test-mass limit.  As Blanchet 
\cite{luc02,luchopkins} has argued,
this suggests that the 3PN approximation may
be quite accurate for comparable-mass systems, without the need for
sophisticated resummation techniques.
At the largest angular velocity considered, 3PN terms contribute less than
one per cent of the total binding energy and angular momentum of the orbit.

Next we consider the effects of tidal and rotational distortions.  
We consider systems of identical bodies ($m_1=m_2$) which are corotating
($\tilde{\omega}_1 = \tilde{\omega}_2 = \Omega$).  For neutron stars, we
adopt the maximum values of the apsidal constants ($k_2 = 3/4$ and $k_3 =
3/8$, see Appendix \ref{clairautsec}), 
and choose two representative values of $q = R_a/m_a$ for neutron star
models with reasonable equations of state, 
namely $q
= 4$ and $q = 6$.  The results, plotted as a fraction of the Newtonian
orbital terms,
are shown in Figures 2 and 3, along with the PN
contributions for comparison.   As expected, tidal effects are very
sensitive to the stellar radii.  For $q=4$, the $l=2$ tidal terms become
comparable to the 2PN and 1PN terms only around $m\Omega \sim 0.09$, while
the $l=3$ terms are an order of magnitude smaller.  For $q = 6$, the
$l=2$ tidal terms exceed the 1PN terms already by $m\Omega \sim 0.05$, while
the $l=3$ terms are small, approaching the 2PN terms only at the largest
allowed
$m\Omega \sim 0.07$, corresponding to the point
at which these larger stars are touching.  For irrotational stars
($\tilde{\omega}_1 = \tilde{\omega}_2 = 0$), the tidal effects are very
similar.

\begin{figure}
\leavevmode
\psfig{figure=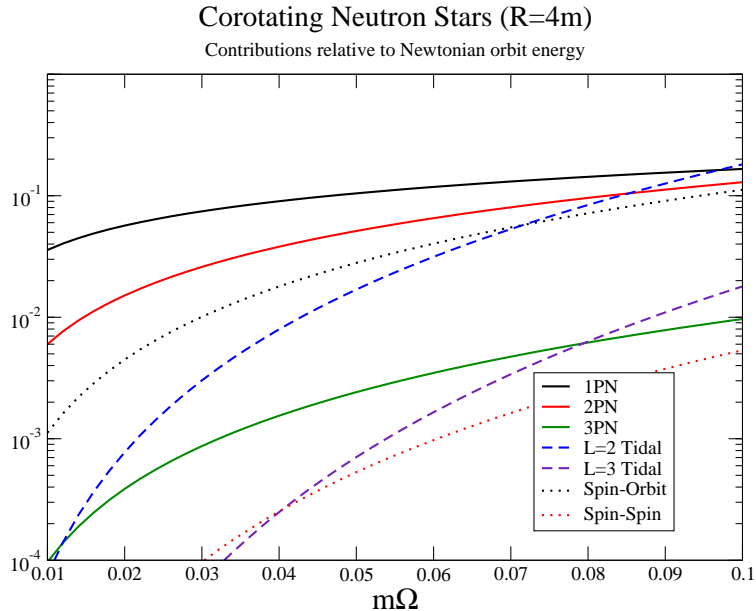,angle=270,width=12cm}
\caption{\label{Fig2} Contributions of tidal and spin terms to the energy
for corotating neutron star binaries,
expressed as a fraction of the Newtonian energy, vs. $m\Omega$, for 
$q=R_a/m_a=4$.  Circular orbits are assumed. 
Shown are the PN contributions for comparison}
\end{figure}

\begin{figure}
\leavevmode
\psfig{figure=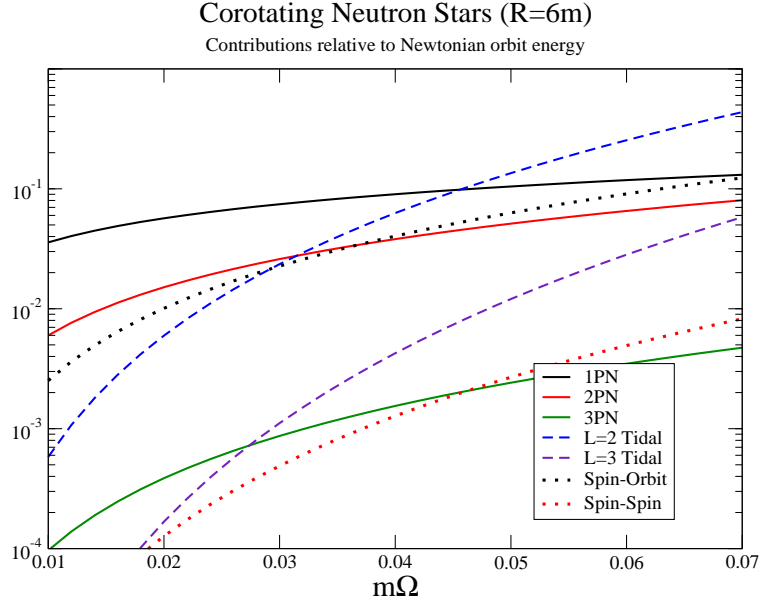,angle=270,width=12cm}
\caption{\label{Fig3} Same as Fig. \ref{Fig2}, with $q=6$}
\end{figure}

These curves illustrate that tidal effects need to be taken into account 
carefully in
an accurate diagnostic for neutron star binaries, 
but are not so large that they invalidate our
approximation scheme.  Their modest size also supports our use of Newtonian
theory to calculate them.  They only become problematical for the largest
neutron stars near the very endpoint of their inspiral.
It should also be pointed out that, in making these estimates, we have adopted
the largest values of the apsidal constants, corresponding to uniform-density
stars.  While neutron stars are not as centrally condensed as, say,
non-degenerate stars, they are also not uniform density, so the $k_l$ may 
well be smaller than their maximum values.  
For example, for a Newtonian polytrope, $p=k\rho^\Gamma$, with
$\Gamma=2$, $k_2=0.26$, so the $q=6$ tidal terms in Fig. \ref{Fig3}
are reduced by a factor of three, bringing them to a level at or below
the 1PN terms over the whole range of $m\Omega$.
On the other hand, 
very little, if anything, is known about the values of $k_l$ for
{\it general relativistic} 
neutron stars over a range of equations of state.  This is a subject that we
are currently investigating.

Figure 4 shows the effects of tides for co-rotating black-hole binaries.  
There we choose $q = 1$ ($R=m$ in harmonic coordinates), $k_2=3/4$ and
$k_3=3/8$ (for slowly rotating black holes, $k_2$ from rotational
distortions
happens to be precisely
$3/4$; see, eg. \cite{hartlethorne}).
We see, not surprisingly, that tidal effects are utterly negligible over the
entire range of $m\Omega$.

\begin{figure}
\leavevmode
\psfig{figure=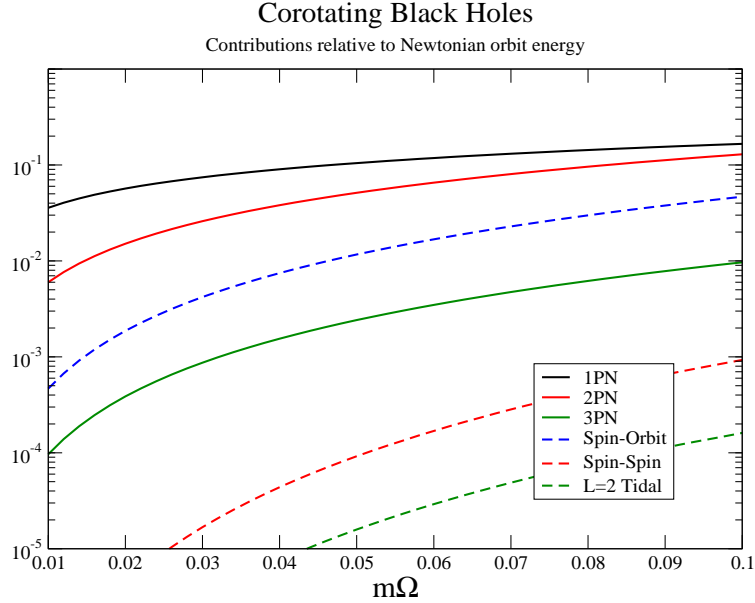,angle=270,width=12cm}
\caption{\label{Fig4} Same as Fig \ref{Fig2} but for corotating
black-hole binaries,
with $q=1$.}
\end{figure}

Finally, we examine spin effects.  Again we consider identical, co-rotating
bodies.  For neutron stars, we assume that $S_a = I_a \Omega$, with the
moment of inertia given by that for a uniform density body, $I_a = (2/5) m_a
R_a^2 = (2/5)q^2 m_a^3$.  The results are shown also in Figures 2 and 3.  For
$q=4$, spin-orbit effects are small but significant, just below the 2PN
terms, while spin-spin effects are negligible.  For $q=6$, spin-orbit
terms exceed 2PN terms by $m\Omega \sim 0.04$ and become comparable to the
1PN terms by the maximum angular velocity, while spin-spin terms barely
exceed the 3PN effects.  

For black holes, we use the fact that $S_a = 4 m_a^3 \Omega$.  Figure 4
shows that the spin-orbit terms lie between the 2PN and 3PN contributions
and thus must be included, while spin-spin terms are negligible (though
larger than the tidal terms). 

\subsection{Corotating, identical black holes}

For black hole binaries, we ignore tidal and spin-spin effects.  We set
$m_1=m_2$, $\eta=1/4$, and $\tilde{\omega}_1=\tilde{\omega}_2 =\Omega$.  
We exploit the fact that there exist
exact formulae for the energy and spin of isolated Kerr black holes in terms
of the irreducible mass, $M=M_{\rm irr}/[1-4(M_{\rm irr}\omega)^2]^{1/2}$,
$S=4M_{\rm irr}^3 \omega/[1-4(M_{\rm irr}\omega)^2]^{1/2}$.  The total
energy and angular momentum of the system are then given by
\begin{eqnarray}
E_{\rm Tot} &=& E_{\rm Self} + E_{\rm Harm} + E_{\rm N,Corr} + E_{\rm Spin} \,,
\nonumber \\
J_{\rm Tot} &=& S + J_{\rm Harm} + J_{\rm N,Corr} + J_{\rm Spin} \,,
\end{eqnarray}
where
\begin{subequations}
\begin{eqnarray}
E_{\rm Self} &=& m_{\rm irr} \left [ 1+\frac{1}{2}(m_{\rm irr}\Omega)^2
+ \frac{3}{8}(m_{\rm irr}\Omega)^4 + \dots \right ]
\,, 
\label{Ebhself}
\\
S &=& m_{\rm irr}^3\Omega \left [ 1 + \frac{1}{2}(m_{\rm irr}\Omega)^2
+ \frac{3}{8}(m_{\rm irr}\Omega)^4 + \dots \right ]
\,, 
\label{Jbhself}
\\
E_{\rm Harm} &=&
-\frac{1}{8}m_{\rm irr} (1-e^2)\zeta \left [ 1 - \frac{1}{48} (37-e^2)\zeta
-\frac{1}{384}(1069-934e^2+129e^4)\zeta^2
\right .
\nonumber \\
&& \left . +\left ( \frac{1}{331776} (1427365+18249e^2-6225e^4-23005e^6) 
- \frac{41 \pi^2}{384} (5-e^2)  \right ) \zeta^3 \right ] \,,
\label{Eharmdiag}
\\
J_{\rm Harm} &=&
\frac{1}{4}m_{\rm irr}^2 \frac{1}{\sqrt{\zeta}}
\left [ 1 + \frac{1}{24}(37-e^2)\zeta
+\frac{1}{384}(1069+234e^2-127e^4) \zeta^2
\right .
\nonumber \\
&& \left .
-\left (\frac{1}{82944}(285473-181851e^2+205683e^4+2311e^6)
-\frac{41 \pi^2}{96}(1+e^2) \right ) \zeta^3 \right ]
\,,
\label{Jharmdiag}
\\
E_{\rm N,Corr} &=& -\frac{5}{48}m_{\rm irr}(1-e^2)(m_{\rm irr}\Omega)^2
\zeta \,,
\\
J_{\rm N,Corr} &=& \frac{5}{24}m_{\rm irr}^2 (m_{\rm irr}\Omega)^2
/\sqrt{\zeta} \,,
\\
E_{\rm Spin} &=& -\frac{1}{12} m_{\rm irr}(1-e^2)(7-2e^2)(m_{\rm irr}\Omega)
\zeta^{5/2} \,,
\\
J_{\rm Spin} &=& -\frac{5}{24}m_{\rm irr}^2 (7+e^2) (m_{\rm irr}\Omega) \zeta \,,
\end{eqnarray}
\end{subequations}
where $m_{\rm irr}$ is the total irreducible mass of the system, given by 
$(m_{\rm irr})_1 + (m_{\rm irr})_2$.  
In Eqs. (\ref{Ebhself}) and (\ref{Jbhself}), we
have expanded
the Kerr formulae for $M$ and $S$ in powers of $m_{\rm irr}\Omega$, assumed to
be
small
compared to unity, keeping as many higher-order terms as needed 
to reach a precision
comparable to our 3PN formulae.
To obtain $E_{\rm Tot}$ and $J_{\rm Tot}$ at a turning point as functions of
$\Omega$, we substitute $\zeta = (m_{\rm irr}\Omega_a)^{2/3}/(1-e)^{4/3}$
or $\zeta = (m_{\rm irr}\Omega_p)^{2/3}/(1+e)^{4/3}$ for apocenter or
pericenter, respectively (in calculating $E_{\rm N,Corr}$ and $J_{\rm
N,Corr}$, we have already changed the dependence in $\zeta$ from the total
mass
of the rotating bodies to the total irreducible
mass of the non-rotating counterparts).  
These are the formulas used in \cite{morawill} to
compare with the numerical HKV quasi-equilibrium solutions of Grandcl\'ement
{\it et al.} \cite{ggb02}.  When $E_{\rm Tot}$ and $J_{\rm Tot}$ are scaled
by $m_{\rm irr}$ and $m_{\rm irr}^2$ respectively, there remains only one
free parameter, the eccentricity of the orbit, and we found \cite{morawill}
that a substantially
better fit to the numerical data was obtained for non-zero values of
$e$, of the order of $0.03$, with the system at apocenter, 
than for $e=0$.  We suggested that such
apparent eccentricity could be a result of the inevitable approximations
(such as the conformally flat approximation) and numerical errors in such
initial-data models, but, in the absence of detailed estimates of the sizes of
those errors, it was difficult to draw firm conclusions.  On the other hand,
those engaged in numerical models of black hole binaries
could use our diagnostic
as a guide to know when, say, a suitable circular orbit has been achieved, or
whether further numerical experiments with different grid sizes or larger
computational domains are necessary to reach the desired physically
meaningful state.

\subsection{Corotating, identical neutron stars}

For neutron stars, we must include tidal effects.  
We set 
$m_1=m_2$, $\eta=1/4$, and $\tilde{\omega}_1=\tilde{\omega}_2 =\Omega$; we let
the apsidal constants and radius factors be
common for both stars, given by $k_2$, $k_3$, and $q$, respectively, and 
express all quantities in terms of the total mass $m_0 = (m_0)_1 +
(m_0)_2$ of two non-rotating stars with the same equation of state.  We also
define for each star the coefficient $\alpha_a = I_a/m_aR_a^2$, and also
assume it to be common for both stars.  The result is
\begin{eqnarray}
E_{\rm Tot} &=& E_{\rm Self} + E_{\rm Harm} + E_{\rm N,Corr} + E_{\rm Spin}
+ E_{\rm TR,Orbit} + E_{\rm Distort} 
\,,
\nonumber \\
J_{\rm Tot} &=& S + J_{\rm Harm} + J_{\rm N,Corr} + J_{\rm Spin} 
+ J_{\rm TR,Orbit} + J_{\rm Distort} \,,
\end{eqnarray}
where the 3PN point-mass expressions $E_{\rm Harm}$ and $J_{\rm Harm}$ are
given in Eqs. (\ref{Eharmdiag}) and (\ref{Jharmdiag}), and where
\begin{subequations}
\begin{eqnarray}
E_{\rm Self} &=& m_0 + \frac{1}{8} \alpha q^2 m_0 (m_0\Omega)^2 \,,
\\
S &=& \frac{1}{4} \alpha q^2 m_0^2 (m_0\Omega)
\,, \\
E_{\rm N,Corr} &=& -\frac{5}{192}m_0(1-e^2)\alpha q^2 (m_0\Omega)^2
\zeta \,,
\\
J_{\rm N,Corr} &=& \frac{5}{96}m_0^2 \alpha q^2 (m_0\Omega)^2
/\sqrt{\zeta} \,,
\\
E_{\rm Spin} &=& -\frac{1}{48} m_0(1-e^2)(7-2e^2)\alpha q^2 (m_0\Omega)
\zeta^{5/2} \,,
\\
J_{\rm Spin} &=& -\frac{5}{96}m_0^2 (7+e^2) \alpha q^2 (m_0\Omega) \zeta
\,,
\\
E_{\rm TR,Orbit}  &=& \frac{1}{32} m_0 (1-e^2) 
\left [ \frac{1}{9} (3-e^2)q^5k_2 (m_0\Omega)^2 \zeta^3
+ \frac{1}{6} (9+10e^2-3e^4)q^5k_2\zeta^6
\right .
\nonumber \\
&& \left .
+\frac{1}{24} (13+49e^2+7e^4-5e^6) q^7 k_3 \zeta^8 \right ]
 \,,
 \\
J_{\rm TR,Orbit}  &=& \frac{1}{32} m_0^2
\left [ \frac{2}{9} (3+e^2)q^5k_2 (m_0\Omega)^2 \zeta^{3/2}
+\frac{2}{3} (3+10e^2+3e^4)q^5k_2\zeta^{9/2}
\right .
\nonumber \\
&& \left .
+\frac{2}{3} (1+7e^2+7e^4+e^6)q^7 k_3 \zeta^{13/2} \right ]
  \,,
   \\
E_{\rm Distort} &=& \frac{1}{48} m_0q^5k_2(m_0\Omega)^2
[(m_0\Omega)^2+(1-\epsilon e)^3 \zeta^3 ]
  \,,
   \\
J_{\rm Distort} &=& \frac{1}{144} m_0^2 q^5k_2(m_0\Omega)
[4(m_0\Omega)^2+3(1-\epsilon e)^3 \zeta^3 ]
      \,.
\end{eqnarray}
\end{subequations}
We illustrate the use of this diagnostic by comparing with numerical data
recently reported by Miller {\it et al.} \cite{mgs03}.  They constructed a
sequence of general relativistic,
quasi-equilibrium configurations of corotating neutron stars, in
the conformally flat approximation.  They used a polytropic equation of
state with $\Gamma=2$.  Among other quantities, they report an ``effective''
binding energy, given by $E_b = [M_{\rm ADM}-2M_{\rm NS}(\Omega)]
/M_0$, as a function of $M_0\Omega$,  
where $M_{\rm ADM}$ is the total ADM mass
of the configuration, 
and $M_{\rm NS}(\Omega)$
is the ADM mass of a uniformly rotating isolated
neutron star of the same baryonic mass $M_0$, as each star in the binary
configuration, but rotating with angular velocity $\Omega$.  
Since the rotational kinetic energy of the stars is already removed, we
can compare the numerical results with the PN diagnostic 
$E_{\rm Diag} = [E_{\rm Tot} - E_{\rm Self}]/m_0$.  Since our $m_0$ is twice the
ADM mass of a non-rotating neutron star, we must scale $E_b$ by $M_0/2M_{\rm
ADM-NS}$, where $M_{\rm
ADM-NS}$ is the ADM mass of an isolated, non-rotating neutron star.  In the
models of Miller {\it et al.}, $M_{\rm
ADM-NS}= M_0 / 1.067$.  We also need to fix the coefficients $q$ and
$\alpha$.  From data provided by Miller (private communication), 
the radius of each isolated
non-rotating star in isotropic coordinates is given by $R_I = 6.77 M_{\rm ADM}$,while the baryonic moment of inertia, calculated using isotropic coordinates,
is given by $I_0 = 9.412 M_0^3$.  We work in harmonic coordinates, but since
$R_H = R_I (1+ M_{\rm ADM}^2/4R_I^2)$, the difference 
between the two coordinates
is only of order 1/2 \%, so we read off $q=6.77$.  The ADM moment
of inertia can be identified as $I_{\rm ADM} = (M_{\rm ADM}/M_0)I_0 = 9.412
M_{\rm ADM}M_0^2 = 9.412 (1.067)^2 M_{\rm ADM}^3$.  Thus we can read
off
$\alpha q^2 =  9.412 (1.067)^2$ and hence $\alpha = 0.234$, or around
half of the uniform-density value of 2/5.  (Miller also
calculates the same quantities in terms of circumferential, or Schwarzschild
radius; after transforming to harmonic coordinates,
the results for $q$ and $\alpha$ 
are consistent with these to within a few per cent.)

\begin{figure}
\leavevmode
\psfig{figure=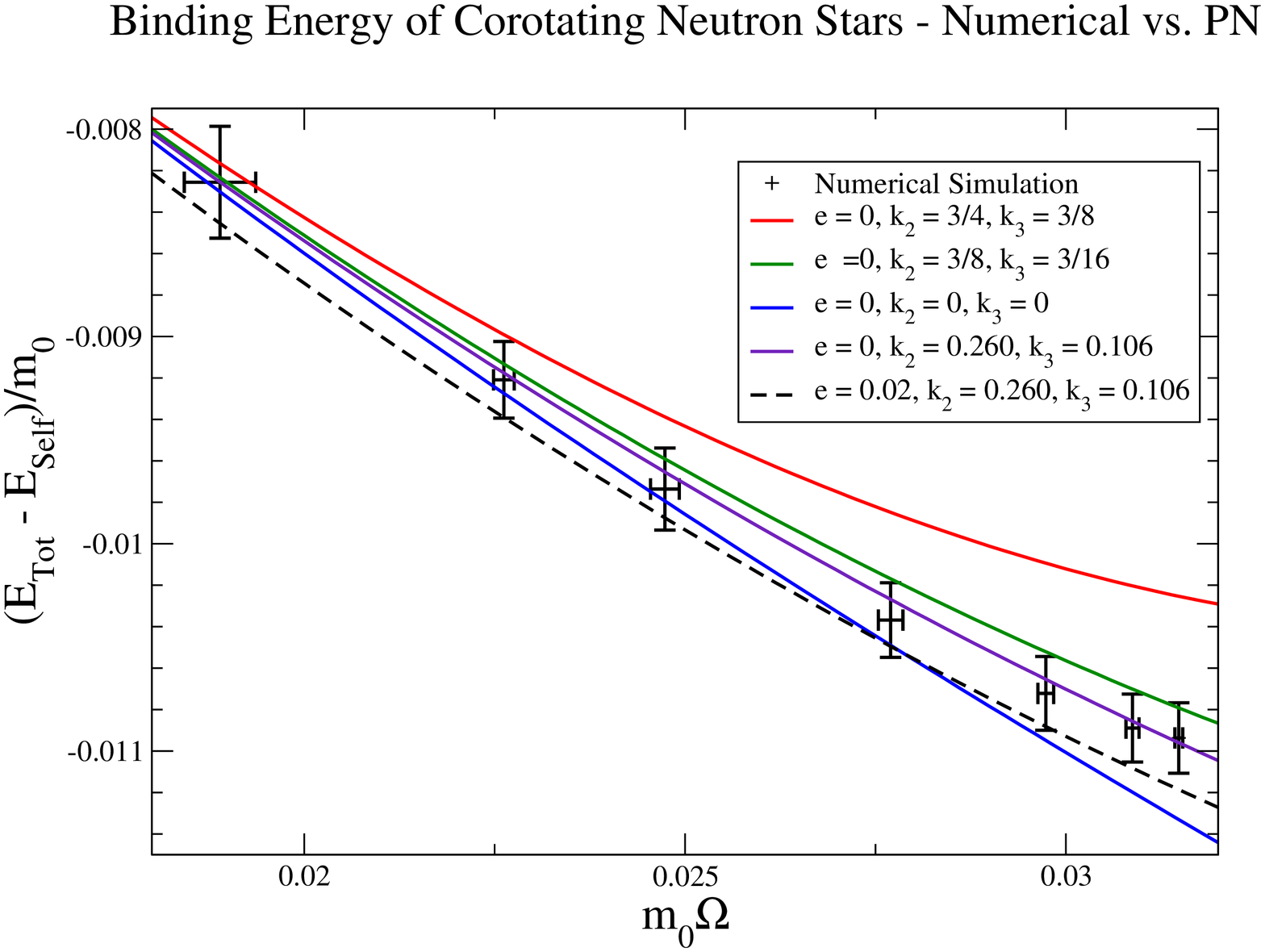,angle=270,width=12cm}
\caption{\label{Fig5} Comparison of PN diagnostic with numerical initial-data
models
of \cite{mgs03}.  }
\end{figure}

Inserting these values of $q$ and $\alpha$ into our diagnostic, we compare
various PN configurations with those reported by Miller {\it et al.}, as
shown in Figure 5.  The numerical results are shown as ``$+$''
with error bars, estimated by Miller {\it et al.} from the results of 
a range of convergence tests.  Four curves
show the energy for 
circular orbits, for various values of the apsidal constants.  Neither
the uniform-density values ($k_2=3/4$, $k_3=3/8$), nor the point mass values
($k_2=0$, $k_3=0$) gives a good fit at all, except at low angular velocities
(large separations) where tidal effects are smaller, and all circular-orbit
curves converge toward the numerical result.  Models with half the
uniform-density values for $k_2$ and $k_3$
give marginal fits.  However, a very good fit is
achieved with values $k_2=0.260$ and $k_3=0.106$; these are
precisely the values for Newtonian
$\Gamma=2$ polytropes (Appendix \ref{appb}), which is the equation of state
used in the Miller {\it et al.} numerical models.  Also shown is
a model with the same $\Gamma=2$  apsidal constants, but with a non-zero
eccentricity $e=0.02$ and with the system at apocenter.  This marginally
fits the numerical data within the error bars, 
but consistently gives lower (more negative) energies.  

We conclude
that these quasi-equilibrium neutron-star configurations 
are fit to better than one per cent by our PN
diagnostic with a circular orbit, and with physically reasonable tidal
terms.

In future work we plan to compare this diagnostic with results of other
numerical models of quasi-equilibrium 
black hole and neutron star binaries.  Our 3PN equations of motion, together
with tidal and spin terms, augmented by radiation reaction terms, can also
be used to develop a ``dynamical'' diagnostic, to compare with numerical
simulations of evolutions from the quasi-equilibrium initial data
\cite{mgs03,mdsb03}.

\begin{acknowledgments}

We are grateful for useful discussions with Emanuele Berti,
Luc Blanchet, Eric Gourgoulhon,
Sai Iyer,
Mark Miller, and
Wai-Mo Suen.
This work was supported in part by the National Science Foundation under
grant No. PHY 00-96522.  T.M. was supported in part by an internship from 
the \'Ecole Normale Sup\'erieure.  C.W. thanks the Institut d'Astrophysique de
Paris for its hospitality during the academic year 2003-2004.
\end{acknowledgments}

\appendix

\section{Newtonian Tidal and Rotational Effects}
\label{appa}

\subsection{Distorted equilibrium configurations}

To derive the effects of tidal and rotational flattening, we will
adopt standard methods from Newtonian theory for binary systems, 
such as those detailed by
Kopal \cite{kopal1,kopal2}.  We assume that the timescale for changes
in perturbing quantities (such as the external tidal potential, seen
either from the global inertial frame, or from the rotating frame of a
given body) is sufficiently long that each body can be assumed to be
in hydrostatic equilibrium.  In other words, we will ignore {\it
dynamical} tides \cite{dynamicaltides}.  This is a 
reasonable assumption as long as we are focusing on quasi-equilibrium
initial data.
Consider one of the bodies in the binary system.
From the equation of hydrostatic equilibrium,
$\nabla p = \rho \nabla \Psi$, where $p$, $\rho$ and $\Psi$ are the
pressure, density and total gravitational potential, respectively, we
conclude that $\nabla \rho \times \nabla \Psi = 0$, and thus that
surfaces of constant $\rho$ and $\Psi$ coincide.  We label surfaces
of constant 
$\rho$ by the radial parameter $a$, and  let the equation of those
surfaces have the form
\begin{equation}
r(a,\theta,\phi) = a[1+ \sum_{l,m} f_{lm}(a) Y_{lm}(\hat \Omega ) ]
\,,
\label{surfaces}
\end{equation}
where $Y_{lm}(\hat \Omega )$ are spherical harmonics corresponding to the
direction ${\hat \Omega}$, and where the dimensionless distortion functions
$f_{lm}$ have the property $f_{lm}^* = (-1)^m f_{l,-m}$.  

On general grounds we expect $f_{lm} \sim (R/r)^{l+1} \sim 
q^{l+1}(m/r)^{l+1}$ for
tidal effects, and, for $l=2$, $f_{2m} \sim \omega^2/\rho \sim (R/r)^3 \sim
q^3 (m/r)^3$ from rotational effects.  The effect of 
these distortions on the
external potential of a body is of order $f_{lm} (R/r)^l \sim 
q^{2l+1}(m/r)^{2l+1}$.
For $l=2$, this means effectively 5PN order; $l=3$ effects would be
effectively 7PN order, and so on.  However, for neutron stars,
with $q \sim 4$ and $m/r
\sim 0.1$, an $l=2$ distortion effect becomes numerically comparable to a
2PN term, while $l=3$ is comparable to a 3PN term.  For black holes, with
$q < 1$, the effects are much smaller.
Thus, in the end, we will keep only $l=2$ and $l=3$ distortion
terms.  Also, non linear corrections to $f_{lm}$ would be of order $(R/r)^{l+1}
\sim q^{l+1} (m/r)^{l+1}$ 
smaller than the dominant linear effects, and thus,
effectively of 8PN order for $l=2$ (for neutron stars, these non-linear
corrections would be numerically smaller than 3PN).  
The exception to this is in the internal
gravitational energy of each body, where a quadratic contribution yields
$(m^2/R){f_{lm}}^2 \sim (m^2/r)(R/r)^{2l+1}$, which is comparable to the
other effectively 5PN contributions for $l=2$.

We begin, however, 
with a general analysis, keeping $lm$ arbitrary, and working to
second order in 
the small
quantities $f_{lm}$.   Later (Appendix \ref{appb}) we will specialize to $l=2$
and $l=3$ linear perturbations.  To second order, 
it is straightforward to show that, for any $n$,
\begin{equation}
r^n = a^n \{1 + n\sum_{l,m} [f_{lm}(a)+(n-1)X_{lm}] Y_{lm}(\hat \Omega )\}
\,,
\label{surfaces2}
\end{equation}
where
\begin{equation}
X_{lm} \equiv \frac{1}{2} \sum_{\alpha\beta;\gamma\delta}
C^{lm}_{\alpha\beta;\gamma\delta} f_{\alpha\beta} f_{\gamma\delta}
\,,
\label{X}
\end{equation}
and $C^{lm}_{\alpha\beta;\gamma\delta}$ is defined in terms of
Clebsch-Gordan coefficients,
\begin{equation}
C^{lm}_{\alpha\beta;\gamma\delta} = \sqrt{
\frac{(2\alpha+1)(2\gamma+1)}{4\pi(2l+1)}} 
\left ( \begin{array}{ccc}
\alpha & \gamma & l \\ 0&0&0
\end{array} \right )
\left ( \begin{array}{ccc}
\alpha & \gamma & l \\ \beta&\delta&m
\end{array} \right )
\,.
\end{equation}
Note that the various angular momentum quantum numbers are connected
by the constraints $l=\alpha + \gamma \,, \alpha + \gamma -2 \,, \dots
\,, |\alpha -\gamma|$, and $m=\beta+\delta$; the
$C^{lm}_{\alpha\beta;\gamma\delta}$ are symmetric under $(\alpha\beta)
\rightleftharpoons (\gamma\delta)$.  Also note that $X_{00} =
(16\pi)^{-1/2} \sum_{\alpha\beta} f_{\alpha\beta}f_{\alpha\beta}^*$.

We expand the gravitational potential $U$ of the body and the
disturbing potential $V$ in the form
\begin{eqnarray}
U &=& \sum_{lm} \frac{4\pi}{2l+1} \int \rho({\bf x}^\prime)
\frac{r_<^l}{r_>^{l+1}} Y^*_{lm}({\hat \Omega}_<) Y_{lm}({\hat \Omega}_>) d^3x^\prime
\,,
\nonumber \\
V &=& 
\tilde{d} r^2 + \sum_{lm} \frac{4\pi}{2l+1} d_{lm} r^l Y_{lm}({\hat \Omega}) 
\,,
\label{UV}
\end{eqnarray}
where the subscript $>(<)$ corresponds to the larger (smaller) of $r$
and $r^\prime$.  The disturbing potential consists of a part, with
disturbing coefficients $d_{lm}$, that
corresponds to a potential with $\nabla^2 V =0$, such as the
gravitational potential from another body, or the Laplacian-free part of a
centrifugal potential, plus the spherical part of a centrifugal
potential, with coefficient $\tilde{d}$.  
We now substitute Eqs. (\ref{surfaces})
and (\ref{surfaces2}) into (\ref{UV}), convert all expressions
from $r$ to  $a$, and
demand that, for $l=0$, the external gravitational potential of our
body have the form $U=m/r$ (i.e. the perturbation does not change the mass
of the body), and that, for $l \ne 0$, the total potential
$U+V$ be constant at a given $a$.  The first can be satisfied if
$f_{00} + 2 X_{00} =0$, while the second holds if, for $l \ne 0$,
\begin{eqnarray}
a^{-l-1} F_{lm}(a) &+& a^l E_{lm}(a) - \frac{2l+1}{4\pi}\frac{m(a)}{a}
f_{lm}(a) + a^l d_{lm} 
\nonumber \\
&=& {\sum_{\alpha\beta;\gamma\delta}}^\prime
 \frac{2l+1}{2\alpha+1} C^{lm}_{\alpha\beta;\gamma\delta}
f_{\gamma\delta}
 \biggl [(\alpha+1)a^{-\alpha-1}F_{\alpha\beta}(a)
  -\alpha a^\alpha
(E_{\alpha\beta}(a)+d_{\alpha\beta}) \biggr ]
\nonumber \\
&&
-\frac{2l+1}{2\pi}\frac{m(a)}{a} X_{lm}(a) 
-2(2l+1)\tilde{d} a^2 f_{lm} \,,
\label{clairaut1}
\end{eqnarray}
where
\begin{eqnarray}
m(a) &=&\int_0^a 4\pi \rho(a)a^2da \,,
\nonumber \\
 F_{lm}(a) &=& \int_0^a \rho(a)da \frac{d}{da} \biggl [a^{l+3}(f_{lm}
 +(l+2)X_{lm}) \biggr ] \,,
\nonumber \\
E_{lm}(a) &=& \int_a^A \rho(a)da \frac{d}{da} \biggl [a^{2-l}(f_{lm}
 +(1-l)X_{lm}) \biggr ] \,,
\end{eqnarray}
and $A$ denotes the surface of the body.  The left-hand-side of Eq.
(\ref{clairaut1}) is first order in $f_{lm}$, while the
right-hand-side is second-order.  Dividing the first-order terms by
$a^l$, differentiating with respect to $a$ and multiplying by
$a^{2l+2}$, we obtain the first-order result
\begin{equation}
F_{lm}(a)= \frac{m(a)a^l}{4\pi} [(l+1)f_{lm}-af_{lm}^\prime ] \,,
\end{equation}
where prime denotes differentiation with respect to $a$.   
Substituting this and the first-order solution of Eq.
(\ref{clairaut1})
back into the right-hand-side of Eq.
(\ref{clairaut1}), it is straightforward to show that the term
involving ${\sum^\prime_{\alpha\beta;\gamma\delta}}$ reduces to
$m(a)(2X_{lm} - a X_{lm}^\prime)/4\pi a$, to second order.  
The basic equation for the
distortion functions $f_{lm}$ can then be written in the form of an integral
equation
\begin{eqnarray}
\frac{2l+1}{4\pi}m(a)f_{lm}(a) 
&-& a^{-l} \int_0^a \rho (a^{l+3}f_{lm})^\prime da 
-a^{l+1} \int_a^A \rho (a^{2-l}f_{lm})^\prime da
\nonumber \\
& = & a^{l+1} d_{lm} + {\cal R}_{lm}(a) \,,
\label{clairaut2}
\end{eqnarray}
where ${\cal R}_{lm}$ contains all contributions quadratic in small
quantities:
\begin{eqnarray}
{\cal R}_{lm}(a)  &=& \frac{2l+1}{4\pi}am(a) X_{lm}^\prime +
2(2l+1)\tilde{d} a^2 f_{lm} 
\nonumber \\
&& +(l+2) a^{-l} \int_0^a \rho (a^{l+3}X_{lm})^\prime da 
\nonumber \\
&&
- (l-1) a^{l+1} \int_a^A \rho (a^{2-l}X_{lm})^\prime da
\,.
\label{clairaut2a}
\end{eqnarray}

Combining Eq. (\ref{clairaut2}) with various derivatives of it, one
obtains the following useful equations, evaluated at the surface $a=A$
of the star:
\begin{subequations}
\begin{eqnarray}
(l+1)f_{lm}(A) - Af_{lm}^\prime (A)
 + {\cal P}_{lm}  
&=&
\frac{4\pi}{m} A^{-l} 
\int_0^A \rho [a^{l+3}(f_{lm}+(l+2)X_{lm})]^\prime da 
 \,,
\label{clairaut3a}
\\
l f_{lm}(A) +Af_{lm}^\prime (A)
-{\cal Q}_{lm}
&=& 
\frac{4\pi}{m} A^{l+1} d_{lm} 
\,,
\label{clairaut3b}
\end{eqnarray}
\label{clairaut3}
\end{subequations}
where $m=\int_0^A 4\pi \rho a^2 da$, and  
\begin{eqnarray}
{\cal P}_{lm} &=& A^2 X_{lm}^{\prime\prime}(A) 
- lA X_{lm}^{\prime}(A) + \frac{8\pi}{m}\tilde{d}A^2 [Af_{lm}^{\prime}(A) -
(l-1)f_{lm}(A)]  \,,
\nonumber \\
{\cal Q}_{lm} &=& A^2 X_{lm}^{\prime\prime}(A)
+ (l+1)A X_{lm}^{\prime}(A)]+ \frac{8\pi}{m}\tilde{d}A^2 [Af_{lm}^{\prime}(A)
+ (l+2)f_{lm}(A)]  \,.
\end{eqnarray}

Another combination of first and second 
derivatives of Eq. (\ref{clairaut2}) yields a second-order
differential equation for $f_{lm}$, sometimes called Clairaut's
equation: 
\begin{equation}
a^2 f_{lm}^{\prime\prime} + \frac{8\pi \rho a^3}{m(a)}
(af_{lm}^{\prime} + f_{lm}) - l(l+1)f_{lm} 
 = \frac{4\pi}{m(a)} [a^2 {\cal R}_{lm}^{\prime\prime}
 - l(l+1){\cal R}_{lm} ] \,.
\label{radau}
\end{equation}
For a given density distribution $\rho(a)$, this equation can be
solved, subject to the boundary conditions that $f_{lm}$ be regular at
$a=0$, and that, at the surface, $f_{lm}$
satisfy Eq.
(\ref{clairaut3b}). 

\subsection{Energy and angular momentum of the system}

Given a solution for the distortion functions $f_{lm}(a)$, we can
calculate all the quantities needed for the equations of motion and
the energy and angular momentum of the orbit.  The external potential
of our body, for example, is given by
\begin{eqnarray}
U &=& \frac{m}{r} + {\sum_{lm}}^\prime \frac{4\pi}{2l+1}
 \frac{Y_{lm}({\hat \Omega})}{r^{l+1}} 
  \int_0^A \rho [a^{l+3}(f_{lm}+(l+2)X_{lm})]^\prime da
\nonumber \\
&=& \frac{m}{r} + {\sum_{lm}}^\prime \frac{4\pi}{2l+1}
 \frac{Y_{lm}({\hat \Omega})}{r^{l+1}}
 \frac{mA^l}{4\pi}
 [(l+1)f_{lm}(A) - Af_{lm}^\prime (A)
 + {\cal P}_{lm} ] 
\nonumber \\
&=& \frac{m}{r} + {\sum_{lm}}^\prime \frac{8\pi}{2l+1}
 \frac{A^{2l+1}}{r^{l+1}} k_{lm} d_{lm} Y_{lm}({\hat \Omega}) \,,
 \label{externalpotential}
\end{eqnarray}
where $\sum^\prime$ denotes summation for $l \ne 0$, and
where we define the ``apsidal constant'' $k_{lm}$ by
\begin{equation}
k_{lm} \equiv \frac{1}{2}
\frac{(l+1)f_{lm}(A) - Af_{lm}^\prime (A) + {\cal P}_{lm}}
{l f_{lm}(A) +Af_{lm}^\prime (A)
-{\cal Q}_{lm}} \,.
\label{apsidal}
\end{equation}
The total gravitational energy of the system is given by
\begin{equation}
W = -\frac{1}{2} \int \int \frac{\rho({\bf x})\rho({\bf x}^\prime)}
{|{\bf x}-{\bf x}^\prime|} d^3x d^3x^\prime 
 =  W_{11} + W_{12} + (1 \rightleftharpoons 2) \,.
\end{equation}
The self-energy $W_{11}$ of body 1 can be written
\begin{equation}
W_{11} = - \sum_{lm} \frac{4\pi}{2l+1}
\int_0^R \rho r^{1-l} drd\Omega Y_{lm}({\hat \Omega})
\int_0^r \rho^\prime {r^\prime}^{l+2} dr^\prime d\Omega^\prime
Y_{lm}^*({\hat \Omega}^\prime) \,.
\end{equation}
Substituting Eqs. (\ref{surfaces}) and (\ref{surfaces2})
we find no contribution linear in $f_{lm}$.  To second order, we
obtain
\begin{eqnarray}
W_{11} &=& -\int_0^A \frac{m(a)}{a} dm(a)
\nonumber \\
&&
+ {\sum_{lm}}^\prime \frac{1}{2l+1} \int_0^A \rho(a)m(a)ada 
\biggl [a^2 |{f_{lm}^\prime}|^2 + 2af_{lm}^*{f_{lm}^\prime} 
+ (l^2+l-1)|{f_{lm}}|^2
\biggr ] \,.
\label{W110}
\end{eqnarray}
Since the second term is already second order, we can integrate by
parts and use the
first-order versions of Eqs. (\ref{clairaut3}) and (\ref{radau}) to obtain
the alternative form
\begin{equation}
W_{11} = - {\cal W}
+ {\sum_{lm}}^\prime \frac{4\pi}{2l+1} A^{2l+1} |{d_{lm}}|^2 k_{lm} \,,
\label{W11}
\end{equation} 
where we define the self-gravitational binding energy $\cal W$ of the
undistorted configuration by
\begin{equation}
{\cal W} = \int_0^A \frac{m(a)}{a} dm(a) \,.
\label{selfgravity}
\end{equation}
In $W_{12}$, we substitute the external potential 
of body 2 evaluated inside body 1, to obtain
\begin{equation}
W_{12} = -\frac{1}{2} \int_1 \rho d^3x \left (
\frac{m_2}{y_2} + {\sum_{lm}}^\prime \frac{8\pi}{2l+1}
\frac{A_2^{2l+1}}{y_2^{l+1}} d_{lm}^{(2)} k_{lm}^{(2)} Y_{lm} ({\hat
y}_2) \right ) \,,
\label{interaction}
\end{equation}
where ${\bf y}_2 = {\bf x} - {\bf x}_2$, $ k_{lm}^{(2)}$ is the
apsidal constant of body 2, and $d_{lm}^{(2)}$ is the coefficient of the
disturbing potential acting on body 2.
Since the interaction energy is smaller than the self energy by a
factor of $R/r$, we only need to keep terms linear in the deformations
$f_{lm}$ or the disturbing coefficients $d_{lm}$; 
consequently we carry out a multipole expansion of $1/y_2$
in the first term in Eq. (\ref{interaction}), then convert from $r$ to
$a$ using Eq. (\ref{surfaces}), but we evaluate the
second term at the center of mass of body 1 and do the lowest-order
spherical integral.  Effectively, we are ignoring multipole-multipole
coupling between the bodies, which can be shown to lead to effects of order
$(A/r)^{2l+1}(A/r)^{2n+1} \sim (m/r)^{10}$, or 10PN order for $l=n=2$.  
The result is
\begin{equation}
W_{12} = -\frac{1}{2} \frac{m_1m_2}{r} 
- \sum_{lm} \frac{4\pi}{2l+1} \left ( m_1 A_2^{2l+1} d_{lm}^{(2)}
k_{lm}^{(2)} + (-1)^l m_2A_1^{2l+1} d_{lm}^{(1)}
k_{lm}^{(1)} \right ) \frac{Y_{lm}({n})}{r^{l+1}} \,,
\label{W12}
\end{equation}
where now $r=|{\bf x}_1 -{\bf x}_2|$ and ${\bf n} =({\bf x}_1
-{\bf x}_2)/r$.
Combining Eqs. (\ref{W11}) and (\ref{W12}), the final result is
\begin{eqnarray}
W &=& - {\cal W}_1- \frac{1}{2} \frac{m_1m_2}{r}
\nonumber \\
&& + \sum_{lm} \frac{4\pi}{2l+1} A_1^{2l+1} d_{lm}^{(1)}
k_{lm}^{(1)} \left ( d_{lm}^{*(1)} - 2(-1)^l \frac{m_2}{r^{l+1}} 
Y_{lm}({n}) \right )
\nonumber \\
&& + (1 \rightleftharpoons 2) \,,
\label{Wfinal}
\end{eqnarray}
where, under the interchange, ${\bf n} \to - {\bf n}$.  

The kinetic energy of the system is given by $T=\int \rho v^2 d^3x$.
Splitting the velocity of an element of fluid into center-of-mass,
rotational, and random parts, 
and noting that 
\begin{equation}
\sum_m Y_{lm}^* ({n})Y_{lm}({n}^\prime) = 
\frac{(2l+1)!!}{4\pi l!} {n}^{<L>} {n^\prime}^{<L>}
=
\frac{2l+1}{4\pi} P_l ({\bf n} \cdot {\bf n}^\prime ) \,, 
\label{Yident}
\end{equation}
where ${\hat n}^{<L>}$ denotes an STF product of $l$ unit vectors (a
capitalized superscript denotes a multi-index), the product ${n}^{<L>}
{n^\prime}^{<L>}$ denotes contraction on all indices, and 
$P_l$ is a Legendre polynomial, 
we may write
\begin{equation}
T = \frac{1}{2} m_1 v_1^2 + T_{\rm Thermal}^{(1)} +
 \frac{1}{3} \omega_1^2 \int_0^{R_1} \rho r^2 dr d\Omega 
 \left ( 1- \frac{4\pi}{5}
   \sum_m Y_{2m}({\hat \lambda}_1) Y_{2m}^*({\hat \Omega}) \right ) 
   + (1 \rightleftharpoons 2) \,,
   \end{equation}
where $\fourvec{\hat \lambda}_1 = \fourvec{\omega}_1/\omega_1$.  
Converting from $r$ to $a$
using Eq. (\ref{surfaces}), recalling that $f_{00} = -2X_{00}$, and noting
that $\omega^2$ is already of first order in disturbing quantities, we obtain,
to second order in small quantities,
\begin{equation}
T = \frac{1}{2} m_1 v_1^2 + T_{\rm Thermal}^{(1)} +
 \frac{1}{2} I_1 \omega_1^2 - \frac{8\pi}{15}\omega_1^2 A_1^5
 \sum_m  d_{2m}^{(1)}k_{2m}^{(1)}Y_{2m}({\hat \lambda}_1)
 + (1 \rightleftharpoons 2) \,,
 \label{Tfinal}
\end{equation}
where $I_1 = (2/3) \int_0^{A_1} 4 \pi \rho a^4 da$.

The angular momentum of the system is given by $J^i = \epsilon^{ijk} \int
\rho x^j v^k$.  Using the same split of the velocities, we obtain, to the
analogous order of precision,
\begin{equation}
J^i = m_1 ({\bf x}_1 \times {\bf v}_1 )^i + I_1 \omega_1^i
-2\omega_1^j A_1^5 \sum_m  d_{2m}^{(1)}k_{2m}^{(1)} Z_{2m}^{<ij>} 
+ (1 \rightleftharpoons 2) \,,
\label{Jfinal}
\end{equation}
where we define the symmetric trace free (STF) tensor 
\begin{equation}
Z_{lm}^{<L>} \equiv \int Y_{lm}({\hat \Omega}) {\hat n}^{<L>} d^2 \Omega \,,
\end{equation}
with the following properties
\begin{eqnarray}
\sum_m Y_{lm}^*({\hat \lambda}) Z_{lm}^{<L>} &=& {\hat \lambda}^{<L>} \,,
\nonumber \\
{\hat \lambda}^L Z_{lm}^{<L>} &=& \frac{4\pi l!}{(2l+1)!!} Y_{lm}({\hat
\lambda}) \,.
\end{eqnarray}

\subsection{Equations of motion}

The Newtonian equations of motion for body 1 are given
by 
\begin{equation}
a_1^i = \frac{1}{m_1} \int_1 \rho d^3x \int_2 \rho^\prime \nabla^i 
\frac{1}{|{\bf x}-{\bf x}^\prime|} d^3x^\prime  \,.
\end{equation}
We write ${\bf x} = {\bf x}_1 + \bar{\bf x}$ and 
${\bf x}^\prime = {\bf x}_2 + \bar{\bf x}^\prime$ and expand in a Taylor
series about ${\bf x}_1$ and ${\bf x}_2$.  We define the STF multipole moments
$I_a^{<Q>} = \int_a \rho \bar{x}^{<Q>} d^3x$
with $I_a^0 = m_a$ and $I_a^j = 0$.  Finally, we calculate the relative
acceleration $a^i = a_1^i - a_2^i$.  After some manipulation, we obtain the
general result
\begin{eqnarray}
a^i &=& -\frac{mx^i}{r^3} 
+ m \sum_{l=2}^{\infty} \frac{1}{l!} \left (\frac{I_1^{<L>}}{m_1}
+ (-1)^l \frac{I_2^{<L>}}{m_2} \right ) \nabla^{iL} \left ( \frac{1}{r}
\right )
\nonumber \\
&& +m \sum_{l=4}^{\infty} \sum_{p=2}^{l-2} 
\frac{(-1)^{l-p}}{p! (l-p)!} \left ( \frac{I_1^{<P}}{m_1}
\frac{I_2^{L-P>}}{m_2} \right ) \nabla^{iL} \left ( \frac{1}{r} 
\right )\,,
\label{eomtidal}
\end{eqnarray}
where $m=m_1+m_2$ and 
the products of the multipole moment tensors are to be symmetrized on all
indices and made trace-free.  For our distorted bodies, the STF multipole
moments can be shown to be given by
\begin{equation}
I_1^{<L>} = 2A_1^{2l+1} \sum_m 
d_{lm}^{(1)}k_{lm}^{(1)} Z_{lm}^{<L>} \,.
\label{Ifinal}
\end{equation}
With the coefficients $d_{lm} \sim m/r^{l+1}$, we have that $I^{<L>} \sim
mA^{2l+1}/r^{l+1}$, and therefore the multipole-multipole coupling term in
the equation of motion (\ref{eomtidal})
is of order $(m/r^2)(A/r)^{2q+2}$; since $q \ge 4$, this is 10PN and higher.  
As before, 
we ignore multipole-multipole terms.  


\subsection{Multiple disturbance sources}

We will want to consider both tidal disturbances as well as rotation-induced
disturbances.  To see how this affects our general results, we note that the
non-linear corrections to the Clairaut equations never play a role to
the order of accuracy we require, only the linear functions $f_{lm}$,
satisfying linear differential equations, are
needed in the end.  
Let $f_{lm} = g_{lm} + h_{lm}$, where each disturbance function
satisfies the linearized Clairaut equations (\ref{radau}), with a boundary
condition for each determined by the linearized Eq. (\ref{clairaut3b}).  
From the
structure of the formulae for the external potential $U$, the kinetic energy, 
the angular momentum, and the multipole moments, 
it is clear that the coefficient $d_{lm}
k_{lm}$ can simply be replaced by $d_{lm}^{(g)}k_{lm}^{(g)} + d_{lm}^{(h)}
k_{lm}^{(h)}$, where $d_{lm}^{(\alpha)}$ is the amplitude of the disturbing
function for that disturbance, and $k_{lm}^{(\alpha)}$ is the corresponding
apsidal constant, determined from the linearized Eq. (\ref{apsidal}).  
However because it
has a contribution quadratic in disturbing functions, the gravitational
self-energy $W_{11}$ requires some care.  Returning to the expression 
(\ref{W110}),
substituting $f_{lm} = g_{lm} + h_{lm}$ and carrying out the integrations by
parts, using the linearized
Clairaut equations satisfied separately by $g_{lm}$ and $h_{lm}$,
one can show that the coefficient $|{d_{lm}}|^2 k_{lm}$ must be replaced by the
coefficient $|{d_{lm}^{(g)}}|^2 k_{lm}^{(g)} + |{d_{lm}^{(h)}}|^2 k_{lm}^{(h)} +
k_{lm}^{(g)}d_{lm}^{(g)}d_{lm}^{*(h)}+ k_{lm}^{(h)}d_{lm}^{(h)}d_{lm}^{*(g)}$.  

\section{Rotational and $l=2$, $l=3$ tidal distortions}
\label{appb}

\subsection{Disturbing coefficients and apsidal constants}

We focus on the lowest-order $l=2$ and $l=3$ tidal terms.  
The gravitational potential at a point ${\bf x}^\prime$
in body 1 due to body 2 is given by
\begin{equation}
V = m_2 \sum_{lm} \frac{4\pi}{2l+1} \frac{{r^\prime}^l}{r^{l+1}} (-1)^l 
Y_{lm}^*(n) Y_{lm}({\hat \Omega}^\prime) \,,
\end{equation}
where we ignore the contributions to $V$ from the distortions of body 2
(ignore multipole-multipole coupling).
We thus obtain
the tidal coefficient for body 1,
\begin{equation}
d_{lm}^{T(1)} = m_2 (-1)^l Y_{lm}^* ({n})/r^{l+1} \,,
\end{equation}
with the coefficient for body 2 obtained by interchange.  Working to
linearized order, we can factor out the azimuthal $m$ dependence by defining
$f_{lm}^T = f_l^T Y_{lm}^* ({\hat n})$, then the
Clairaut equation and the outer boundary condition for body 1 take the form
\begin{subequations}
\begin{eqnarray}
a^2 {f_l^T}^{\prime\prime} + \frac{8\pi\rho a^3}{m(a)} (a{f_l^T}^\prime +
f_l^T ) - l(l+1)f_l^T &=& 0  \,,
\label{clairauteq}
\\
A_1{f_l^T}^\prime (A_1) + l f_l^T(A_1) &=& 4\pi (-1)^l \frac{m_2}{m_1} 
\left ( \frac{A_1}{r} \right)^{l+1} \,.
\end{eqnarray}
\end{subequations}
Note that the apsidal constant depends only on $f_l^T$, and is thus
independent of $m$, so
\begin{equation}
k_l^{T(1)} =  \frac{l+1- \eta_l^T(A_1)}{2l+2\eta_l^T(A_1)}  
\,,
\label{apsidal2}
\end{equation}
with a corresponding expression for body 2,
where
\begin{equation}
\eta_l^T \equiv  \frac{d(\ln f_l^T )}{d(\ln a)} \,.
\end{equation}
Note that the overall scale of $f_l^T$ has no effect on the apsidal
constant, to linear order.  

For a uniformly
rotating body, the disturbing potential at a point ${\bf x}^\prime$
is the centrifugal potential
\begin{eqnarray}
V_{\rm Rot} &=& \frac{1}{2} \omega^2 [{r^\prime}^2 - 
(\fourvec{\hat \lambda} \cdot
{\bf x}^\prime )^2 ] 
\nonumber \\
&=& \frac{1}{3} \omega^2{r^\prime}^2 - \frac{4\pi}{15} \omega^2{r^\prime}^2
  \sum_m Y_{2m}^* ({\hat \lambda}) Y_{2m}({\hat \Omega}^\prime) \,.
\end{eqnarray}
Thus we read off the rotational coefficient for body 1
\begin{equation}
{d_{2m}^R}^{(1)} = -\frac{1}{3} \omega_1^2 Y_{2m}^* ({\hat \lambda}_1)
\end{equation}
with the coefficient for body 2 obtained by interchange (the spherical
coefficient $\tilde{d}$ only contributes at second order in small
quantities).  Similarly defining
$f_{2m}^R = f_2^R Y_{2m}^* ({\hat \lambda})$, we find that $f_2^R$ also
satisfies Eq. (\ref{clairauteq}) for $l=2$, but with the boundary condition
\begin{equation}
A_1{f_2^R}^\prime (A_1) + 2 f_2^R(A_1) =  -\frac{4\pi}{3} \frac{\omega_1^3
A_1^3}{m_1} 
\,.
\end{equation}
The rotational apsidal constant is also independent of $m$, and, since the
overall scale of $f_2^R (A_1)$ is irrelevant, 
it is equal to the $l=2$ tidal apsidal constant:
$k_2^{R(1)} = k_2^{T(1)} \equiv k_2^{(1)}$.  
This equality will not hold when non-linear
corrections are included (it will also not hold in general relativity, when
frame-dragging and other relativistic effects are included).  

\subsection{Energy, angular momentum and equations of motion}

Substituting these results for $d_{2m}^T$, 
$d_{3m}^T$, and $d_{2m}^R$, into Eqs. (\ref{Wfinal}),
(\ref{Tfinal}), (\ref{Jfinal}), (\ref{eomtidal}), and (\ref{Ifinal}), 
and making use of Eq. (\ref{Yident}),
we obtain
\begin{subequations}
\begin{eqnarray}
W &=& - \frac{\mu m}{r} + \biggl \{ -{\cal W}_1
+  A_1^5 k_2^{(1)} \left [\frac{1}{9} \omega_1^4 -\frac{m_2^2}{r^6}
\right ] 
- A_1^7 k_3^{(1)}\frac{m_2^2}{r^8}
+ (1 \rightleftharpoons 2) \biggr \} \,,
\label{Wnew}
\\
T &=& \frac{1}{2} \mu v^2 + \biggl \{ T_{\rm Thermal}^{(1)} + \frac{1}{2} I_1
\omega_1^2 
+ \frac{1}{3} A_1^5 \omega_1^2 k_2^{(1)} \left [ 
\frac{2}{3} \omega_1^2 - \frac{m_2}{r^3} [3(\fourvec{\hat
\lambda}_1 \cdot {\bf n})^2 - 1] \right ] 
\nonumber \\
&& 
+ (1 \rightleftharpoons 2) \biggr \} \,,
\label{Tnew}
\\
J^i &=& \mu ({\bf x} \times {\bf v})^i + \biggl \{ I_1 \omega_1^i 
+  \frac{2}{3} A_1^5 k_2^{(1)} \left [ \frac{2}{3} \omega_1^2 \omega_1^i
- \frac{m_2}{r^3} (3n^i \fourvec{\omega}_1 \cdot {\bf n} -\omega_1^i )
\right ]
\nonumber \\
&& 
+ (1 \rightleftharpoons 2) \biggr \} \,,
\label{Jnew}
\\
a^i &=& - \frac{mn^i}{r^2} - \biggl \{
\frac{m}{m_1} A_1^5 k_2^{(1)} \left [ 6 \frac{m_2{n}^i}{r^7}
 + \frac{\omega_1^2}{r^4}[n^i - 5n^i (\fourvec{\hat \lambda}_1 \cdot {\bf n})^2
 + 2 {\hat \lambda}_1^i \fourvec{\hat \lambda}_1\cdot {\bf n} ] \right ]
 \nonumber \\
 && + 8 \frac{m}{m_1} A_1^7  k_3^{(1)} 
 \frac{m_2{n}^i}{r^9} + (1 \rightleftharpoons 2) \biggr \} \,,
 \label{eomnew}
\\
I_1^{<jk>} &=&  2A_1^5 k_2^{(1)} \left ( \frac{m_2}{r^3} {n}^{<jk>} 
- \frac{1}{3} \omega_1^2 {\hat \lambda}_1^{<jk>} \right ) \,,
\\
I_1^{<jkl>} &=&  -2A_1^7 k_3^{(1)} \frac{m_2}{r^3} {n}^{<jkl>} \,.
\end{eqnarray}
\end{subequations}
It is simple to show that the equation of motion (\ref{eomnew})
admits the two conserved
quantities
\begin{subequations}
\begin{eqnarray}
E^* &=& \frac{1}{2} \mu v^2 - \frac{\mu m}{r}
- \biggl \{ A_1^5 k_2^{(1)} \left [ \frac{m_2^2}{r^6} -
\frac{1}{3} \frac{m_2\omega_1^2}{r^3} 
[3(\fourvec{\hat \lambda}_1 \cdot {\bf n})^2 - 1] \right ] 
\nonumber \\
&& +  A_1^7  k_3^{(1)} \frac{m_2^2}{r^8} + (1 \rightleftharpoons 2) \,,
\biggr \}
\label{Estar}
\\
J^{*i} &=& \mu ({\bf x} \times {\bf v})^i + 2m_2 A_1^5 k_2^{(1)}\omega_1^2
B_1^i (t)  + (1 \rightleftharpoons 2) \,,
\label{Jstar}
\end{eqnarray}
\end{subequations}
where $B_1^i (t) = \int^t ({\bf n} \times \fourvec{\hat \lambda}_1)^i
({\bf n} \cdot \fourvec{\hat \lambda}_1) r^{-3} dt$, where we assume that
the various quantities entering the perturbing terms ($A_1$, $k_2^{(1)}$
$\omega_1^i$, $\fourvec{\hat \lambda}_1$ etc.) are constant in time to the order
considered.  By comparing Eqs. (\ref{Jnew}) and (\ref{Jstar}), we see that
the total constant angular momentum can be written in the form
\begin{equation}
J^i = \mu ({\bf x} \times {\bf v})^i + J_1^i +J_2^i + \left \{ 2m_2 A_1^5
k_2^{(1)}\omega_1^2
B_1^i (t)  + (1 \rightleftharpoons 2) \right \}  \,,
\label{Jfinal2}
\end{equation}
where $J_1^i$ and $J_2^i$ are separately constant, defined by
$J_1^i = I_1 \bar{\omega}_1^i$, where the {\em constant} $\bar{\omega}_1^i$
is given by
\begin{equation}
\bar{\omega}_1^i ={\omega}_1^i +  
 \frac{2}{3} \frac{A_1^5 k_2^{(1)}}{I_1} \left [ \frac{2}{3} \omega_1^2 \omega_1^i
 - \frac{m_2}{r^3} (3n^i \fourvec{\omega}_1 \cdot {\bf n} -\omega_1^i )
 -3 m_2 {\omega}_1^2 B_1^i(t)
 \right ] \,.
 \label{baromega}
\end{equation}
Notice that $B_1^i(t)$ is orthogonal to $\bf n$ and
$\fourvec{\lambda}_1$, and vanishes if the body's spin axis is
perpendicular to the orbital plane.  Calculating the total energy $E=T+W$
from Eqs. (\ref{Wnew}) and (\ref{Tnew})
and converting from $\omega_1^i$ to the constant $\bar{\omega}_1^i$, we
obtain the final conserved energy, including tidal and rotational
contributions
\begin{eqnarray}
E &=& \frac{1}{2} \mu v^2 - \frac{\mu m}{r}
+ \biggl \{ 
 \frac{1}{2} I_1 \bar{\omega}_1^2
- {\cal W}_1
- A_1^5 k_2^{(1)} \left [ \frac{1}{9} \bar{\omega}_1^4
+ \frac{m_2^2}{r^6} - \frac{1}{3} \frac{m_2 \bar{\omega}_1^2}{r^3}
[3(\fourvec{\hat \lambda}_1 \cdot {\bf n})^2 - 1] \right ]
\nonumber \\
&& -  A_1^7  k_3^{(1)} \frac{m_2^2}{r^8} + (1 \rightleftharpoons 2) \biggr \} \,.
\label{Efinal2}
\end{eqnarray}
Modulo constants, this is identical to $E^*$, Eq. (\ref{Estar}).  

\subsection{Clairaut's equation and the apsidal constants}
\label{clairautsec}

To determine the tidal and rotational distortion effects in our binary
system, it is sufficient to know the disturbing forces (leading to the
coefficients $d_{lm}$) and the apsidal constants.  To linear order, the
apsidal constants can be obtained from solutions of Eq. (\ref{clairauteq}),
along with Eq. (\ref{apsidal2}); this applies to both tidal and rotational
perturbations.   Because the scale of $f_l$ is
irrelevant to the value of $k_l$, it is useful to 
recast Eq. (\ref{clairauteq}) into a first-order
differential equation for $\eta_l$, sometimes called Radau's equation
\begin{equation}
a {\eta_l}^{\prime} + 6{\cal D} (\eta_l +1) +\eta_l (\eta_l-1) 
-l(l+1) =0 \,,
\label{radau2}
\end{equation}
where ${\cal D} = 4\pi \rho(a) a^3/3m(a) = \rho(a)/\bar{\rho}(a) $, and
where we drop the superscripts $T$ or $R$.
Near the origin, where ${\cal D} \to 1$, 
the regularity of $f_l$ requires that $\eta_l (a) \to
l-2$ as $a \to 0$.  Given a density profile for a spherically symmetric
configuration provided by a chosen equation of state, one integrates Eq.
(\ref{radau2}) from the center to the surface, thereby obtaining $\eta_l(A)$,
and thus $k_l$.  

Exact solutions of Radau's equation are known for special cases.  For a
homogeneous star, with $\rho =$ constant, ${\cal D} \equiv 1$, it is easy to
show that $\eta_l = l-2$, and 
\begin{equation}
{k_l}^{\rm Homogeneous} = \frac{3}{4(l-1)} \,.
\end{equation}
For a point mass, with ${\cal D} = 0$ except at the origin, $\eta_l =
l+1$, and hence, as expected, ${k_l}^{\rm Point} =0$.   Generally, if the
density nowhere increases outwards (i.e. if $\rho^\prime (a) \le 0$), then
$\eta_l (A)$ satisfies the inequalities $l-2 \le \eta_l (A) \le l+1$.
For nearly homogeneous configurations, and for $l=2$, Radau's equation can
be rewritten in the approximate form \cite{kopal1}
\begin{equation}
\frac{d}{da} \left ( \bar{\rho}(a)a^5 \sqrt{1+\eta_2(a)} \right )
= 5\bar{\rho}(a)a^4  \left [ 1 + \frac{1}{40} \eta_2(a)^2 + O(\eta_2(a)^3)
\right ] \,.
\end{equation}
Since $\eta_2(a) = 0$ in the homogeneous limit, one can take the lowest
order approximation, integrate, and then, after some manipulation, show that 
\begin{equation}
\eta_2 \approx 3(1- I/I_H) + O[(1- I/I_H)^2]\,,
\end{equation}
where $I$ is the moment of inertia of the star, and $I_H$ is its homogeneous
counterpart.  Finally, for polytropic Newtonian stars, with equation of
state $p=k\rho^\Gamma = k\rho^{1+1/n}$, Kopal \cite{kopal1} lists computed
values of $\eta_l$ and $k_l$ in 
Tables 2-1 and
2-2, for $l=1 \dots 7$ and $n=0 \dots 5$.  For $n=1$ or $\Gamma=2$, which is
a common choice in numerical models of binary neutron stars, $k_2 =
0.260$ and $k_3 = 0.106$.
Kopal \cite{kopal1,kopal2} explores these and other general properties of
Radau's equation.

It is also useful to be able to read off values of $k_2$ from the external
field  or spacetime geometry of a given solution.
For example, if the perturbing coefficient has the form $d_{lm} = d_{l}
Y_{lm}^*(\hat{\lambda})$, where $\hat{\lambda}$ is a principal axis of
the perturbation, then the external potential 
Eq. (\ref{externalpotential}) takes
the form 
\begin{equation}
U = \frac{m}{r} + 2{\sum_l}^\prime \frac{A^{2l+1}}{r^{l+1}} k_l d_l P_l
(\cos \gamma) \,,
\end{equation}
where $\gamma$ is the angle between $\hat{\lambda}$ and the field point.
Then, if one can read off the multipole moments $Q_l$ from an expansion of
the field in the form $m/r + \sum_l Q_l P_l(\cos \gamma)/r^{l+1}$, then one
can determine the apsidal constants according to
\begin{equation}
k_l = \frac{Q_l}{2A^{2l+1} d_l} \,.
\end{equation}
For $l=2$ rotational perturbations, $d_2 = -\omega^2/3$, so
$k_2=(3/2)Q_2/A^5\omega^2$.  This permits one to determine $k_2$ from a
numerical solution of a rotating neutron star, for example.  For Kerr black
holes, $Q_2=S^2/M$.  To lowest order in $M_{\rm irr}\omega$, where $M_{\rm
irr}$ is the irreducible mass, $S=4M_{\rm irr}^3 \omega$ and
$A=2M_{\rm irr}$, 
so that $k_2^{\rm BH} = 3/4$, precisely the value for a homogeneous
Newtonian star.


\begin{thebibliography}{99}

\bibitem{cook} G. B. Cook, Living. Rev. Relativ. {\bf 3}, 2000-5
(2000) (http://www.livingreviews.org/).

\bibitem{seidel} E. Seidel, in {\it Black Holes and Gravitational
Waves: New Eyes in the 21st Century, Proceedings of the 9th Yukawa
International Seminar, Kyoto, 1999}, edited by T. Nakamura and H.
Kodama [Prog. Theor. Phys. Suppl. {\bf 136}, 87 (1999).

\bibitem{baumshapiro} T. W. Baumgarte and S. L. Shapiro, Phys. Reports {\bf
376}, 41 (2003).

\bibitem{jaraschafer98} P. Jaranowski and G. Sch\"afer, Phys. Rev. D
{\bf 57}, 5948 (1998), {\it ibid}. {\bf 57}, 7274 (1998).

\bibitem{jaraschafer99} P. Jaranowski and G. Sch\"afer, Phys. Rev. D
{\bf 60}, 124003 (1999).

\bibitem{djs00} T. Damour, P. Jaranowski and G. Sch\"afer,  Phys. Rev.
D {\bf 62}, 021501 (2000); Erratum, ibid. {\bf 63}, 029903 (2001).

\bibitem{djs01} T. Damour, P. Jaranowski and G. Sch\"afer,  Phys. Rev.
D {\bf 63}, 044021 (2001); Erratum, ibid. {\bf 66}, 029901 (2002) 

\bibitem{bf00} L. Blanchet and G. Faye,  Phys.Lett. {\bf 271A}, 58
(2000).

\bibitem{bf01} L. Blanchet and G. Faye,  Phys. Rev. D
{\bf 63}, 062005 (2001).

\bibitem{abf01} V. de Andrade, L. Blanchet and G. Faye,  Class.
Quantum Gravit.
{\bf 18}, 753 (2001).

\bibitem{bdgef03} L. Blanchet, T. Damour, and G. Esposito-Farese, Phys.
Rev. D, in press (gr-qc/0311052).

\bibitem{futamase01} Y. Itoh, T. Futamase, and H. Asada,
Phys. Rev. D {\bf 63},
064038 (2001).

\bibitem{itohfuta03} Y. Itoh and T. Futamase, Phys. Rev. D {\bf 68}, 121501
(2003).

\bibitem{dire2} M. E. Pati and C. M. Will, Phys. Rev. D {\bf 65},
104008 (2002).

\bibitem{ibbh} P. R. Brady, J. D. E. Creighton and K. S. Thorne,
Phys.Rev. D {\bf 58}, 061501 (1998).

\bibitem{luc02} L. Blanchet, Phys. Rev. D {\bf 65}, 124009 (2002).

\bibitem{luchopkins} L. Blanchet, in {\it Proceedings of the 25th Johns
Hopkins Workshop}, edited by I. Ciufolini, D. Dominici 
and L. Lusanna (World Scientific,
Singapore, 2003), p. 411.

\bibitem{ggb02} P. Grandcl\'ement, E. Gourgoulhon and S. Bonazzola,
Phys. Rev. D {\bf 65}, 044021 (2002).

\bibitem{miller03} M. Miller, preprint (gr-qc/0305024).

\bibitem{wisemanlai} D. Lai and A. G. Wiseman,
Phys.Rev. D {\bf 54}, 3958 (1996). 

\bibitem{cook94} G. B. Cook, Phys. Rev. D {\bf 50}, 5025 (1994).

\bibitem{pfeiffer} H. P. Pfeiffer, S. A. Teukolsky and G. B. Cook, Phys.
Rev. D {\bf 62}, 104018 (2000).

\bibitem{dgg02} T. Damour, E. Gourgoulhon, P. Grandcl\'ement, Phys.
Rev. D {\bf 66}, 024007 (2002).

\bibitem{morawill} T. Mora and C. M. Will, Phys. Rev. D {\bf 66}, 101501
(2002).

\bibitem{mgs03}  M. Miller, P. Gressman and W.-M. Suen, Phys. Rev. D, {\bf
69}, 064026 (2004).

\bibitem{kopal1} 
Z. Kopal, {\em Close Binary Systems} (Chapman and Hall, London, 1959), Chap.
2.

\bibitem{kopal2} 
Z. Kopal, {\em Dynamics of Close Binary Systems} (D. Reidel, Dordrecht,
1978), Chap. 2.

\bibitem{wagwill} R. V. Wagoner and C. M. Will, Astrophys. J. {\bf
210}, 764 (1976); {\bf 215}, 984 (1977).

\bibitem{lincwill} C. W. Lincoln and C. M. Will, Phys. Rev. D {\bf 42},
1123 (1990).

\bibitem{ds88} T. Damour and G. Sch\"afer, Nuovo Cimento {\bf B101},
127 (1988).

\bibitem{blanchetiyer03} L. Blanchet and B. R. Iyer, Class.
Quantum Gravit.
{\bf 20}, 755 (2003).

\bibitem{djs01lett}
T. Damour, P. Jaranowski and G. Sch\"afer, Phys. Lett. {\bf B513}, 147
(2001)

\bibitem{bender} C. M. Bender and S. A. Orszag, {\em Advanced Mathematical
Methods for Scientists and Engineers} (McGraw Hill, New York, 1978), Chap.
11.

\bibitem{spincomment}  In some contexts, spin-orbit and spin-spin
effects are viewed as effectively 1.5PN and 2PN order, respectively.
In those contexts, the bodies' spins are measured in terms of the
maximum spin $S \sim m^2$ of a Kerr black hole, corresponding to
$\omega \sim 1/m$.  In our context, the bodies will never be more than
corotating, so that spin effects are much smaller.

\bibitem{kww} L. E. Kidder, C. M. Will, and A. G. Wiseman, Phys. Rev.
D {\bf 47}, R4183 (1993).

\bibitem{kidder} L. E. Kidder, Phys. Rev. D {\bf 52}, 821 (1995).

\bibitem{hartlethorne} J. B. Hartle and K. S. Thorne, Astrophys. J. {\bf
153}, 807 (1968).

\bibitem{mdsb03} P. Marronetti, M. D. Duez, S. L. Shapiro and T. W.
Baumgarte, Phys. Rev. Lett. {\bf 92}, 141101 (2004).

\bibitem{dynamicaltides}  For discussion of tidal excitation of normal
modes in neutron star binaries, see 
K. D. Kokkotas and G. Sch\"afer, Mon. Not. R. Astron. Soc. {\bf 275}, 301
(1995), and
W. C. G. Ho and D. Lai, Mon. Not. R. Astron. Soc. {\bf 308}, 153 (1999).  
\end{thebibliography}

\end{document}